\documentclass[nojss]{jss}\usepackage[]{graphicx}\usepackage[]{color}
\makeatletter
\def\maxwidth{ %
  \ifdim\Gin@nat@width>\linewidth
    \linewidth
  \else
    \Gin@nat@width
  \fi
}
\makeatother

\usepackage{Sweave}

\usepackage{amsmath}

\usepackage{xcolor}
\usepackage{booktabs}
\usepackage{floatrow}
\author{John Ehrlinger\\Cleveland Clinic}

\title{\pkg{ggRandomForests}: Random Forests for Regression}

\Plainauthor{Ehrlinger} 
\Plaintitle{ggRandomForests: Random Forests for Regression} 
\Shorttitle{Random Forests for Regression}

\Abstract{ 
Random Forests~\citep{Breiman:2001} (RF) are a non-parametric statistical method requiring no distributional assumptions on covariate relation to the response. RF are a robust, nonlinear technique that optimizes predictive accuracy by fitting an ensemble of trees to stabilize model estimates. The \pkg{randomForestSRC} package~\citep{Ishwaran:RFSRC:2014} is a unified treatment of Breiman's random forests for survival, regression and classification problems.

Predictive accuracy make RF an attractive alternative to parametric models, though complexity and interpretability of the forest hinder wider application of the method. We introduce the \pkg{ggRandomForests} package, tools for visually understand random forest models grown in \proglang{R}~\citep{rcore} with the \pkg{randomForestSRC} package. The \pkg{ggRandomForests} package is structured to extract intermediate data objects from \pkg{randomForestSRC} objects and generates figures using the \pkg{ggplot2}~\citep{Wickham:2009} graphics package.

This document is structured as a tutorial for building random forests for regression with the \pkg{randomForestSRC} package and using the \pkg{ggRandomForests} package for investigating how the forest is constructed. We investigate the Boston Housing data~\citep{Harrison:1978,Belsley:1980}. We demonstrate random forest variable selection using Variable Importance (VIMP)~\citep{Breiman:2001} and Minimal Depth~\citep{Ishwaran:2010}, a property derived from the construction of each tree within the forest. We will also demonstrate the use of variable dependence plots~\citep{Friedman:2000} to aid interpretation RF results. We then examine variable interactions between covariates using a minimal depth interactions, and conditional variable dependence plots. The goal of the exercise is to demonstrate the strength of using Random Forest methods for both prediction and information retrieval in regression settings.
}
\Plainkeywords{random forest, regression, VIMP, minimal depth, R, randomForestSRC}

\Address{
John Ehrlinger\\
Quantitative Health Sciences\\
Lerner Research Institute\\
Cleveland Clinic\\
9500 Euclid Ave\\
Cleveland, Ohio 44195\\
E-mail: \email{john.ehrlinger@gmail.com}\\
URL: \url{http://www.lerner.ccf.org/qhs/people/ehrlinj/}
}

\IfFileExists{upquote.sty}{\usepackage{upquote}}{}
\begin{document}
\section*{About this document}\addcontentsline{toc}{section}{About this document}

This document is a package vignette for the \pkg{ggRandomForests} package for ``Visually Exploring Random Forests'' (\url{http://CRAN.R-project.org/package=ggRandomForests}). The \pkg{ggRandomForests} package is designed for use with the \pkg{randomForestSRC} package~\citep[\url{http://CRAN.R-project.org/package=randomForestSRC}]{Ishwaran:RFSRC:2014} for growing random forests for survival (time to event response), regression (continuous response) and classification (categorical response) settings and uses the \pkg{ggplot2} package~\citep[\url{http://CRAN.R-project.org/package=ggplot2}]{Wickham:2009} for plotting diagnostic and variable association results. \pkg{ggRandomForests} is  structured to extract data objects from \pkg{randomForestSRC} objects and provides functions for printing and plotting these objects.

The vignette is a tutorial for using the \pkg{ggRandomForests} package with the \pkg{randomForestSRC} package for building and post-processing random forests for regression settings. In this tutorial, we explore a random forest for regression model constructed for the Boston housing data set~\citep{Harrison:1978,Belsley:1980}, available in the \pkg{MASS} package~\citep{mass:2002}. We grow a random forest and demonstrate how \pkg{ggRandomForests} can be used when determining how the response depends on predictive variables within the model. The tutorial demonstrates the design and usage of many of \pkg{ggRandomForests} functions and features and also how to modify and customize the resulting \code{ggplot} graphic objects along the way.

The vignette is written in \LaTeX using the \pkg{knitr} package~\citep[\url{http://CRAN.R-project.org/package=knitr}]{Xie:2015, Xie:2014,Xie:2013}, which facilitates weaving \proglang{R}~\citep{rcore} code, results and figures into document text. Throughout this document, \proglang{R} code will be displayed in \emph{code blocks} as shown below. This code block loads the \proglang{R} packages required to run the source code listed in code blocks throughout the remainder of this document.
\begin{Schunk}
\begin{Sinput}
R> ################## Load packages ##################
R> library("ggplot2")         # Graphics engine
R> library("RColorBrewer")    # Nice color palettes
R> library("plot3D")          # for 3d surfaces. 
R> library("dplyr")           # Better data manipulations
R> library("parallel")        # mclapply for multicore processing
R> 
R> # Analysis packages.
R> library("randomForestSRC") # random forests for survival, regression and 
R>                            # classification
R> library("ggRandomForests") # ggplot2 random forest figures (This!)
R> 
R> ################ Default Settings ##################
R> theme_set(theme_bw())     # A ggplot2 theme with white background
R> 
R> ## Set open circle for censored, and x for events 
R> event.marks <- c(1, 4)
R> event.labels <- c(FALSE, TRUE)
R> 
R> ## We want red for death events, so reorder this set.
R> strCol <- brewer.pal(3, "Set1")[c(2,1,3)]
\end{Sinput}
\end{Schunk}

This vignette is available within the \pkg{ggRandomForests} package on the Comprehensive R Archive Network (CRAN)~\citep[\url{http://cran.r-project.org}]{rcore}. Once the package has been installed, the vignette can be viewed directly from within \proglang{R} with the following command:
\begin{Schunk}
\begin{Sinput}
R> vignette("randomForestSRC-Regression", package = "ggRandomForests")
\end{Sinput}
\end{Schunk}

A development version of the \pkg{ggRandomForests} package is also available on GitHub (\url{https://github.com}). We invite comments, feature requests and bug reports for this package at \url{https://github.com/ehrlinger/ggRandomForests}.

\section{Introduction} \label{S:introduction}

Random Forests~\citep{Breiman:2001} (RF) are a fully non-parametric statistical method which requires no distributional or functional assumptions on covariate relation to the response. RF is a robust, nonlinear technique that optimizes predictive accuracy by fitting an ensemble of trees to stabilize model estimates. Random Survival Forests (RSF)~\citep{Ishwaran:2007a,Ishwaran:2008} are an extension of Breiman's RF techniques to survival settings, allowing efficient non-parametric analysis of time to event data. The \pkg{randomForestSRC} package~\citep{Ishwaran:RFSRC:2014} is a unified treatment of Breiman's random forests for survival (time to event response), regression (continuous response) and classification (categorical response) problems.

Predictive accuracy make RF an attractive alternative to parametric models, though complexity and interpretability of the forest hinder wider application of the method. We introduce the \pkg{ggRandomForests} package for visually exploring random forest models. The \pkg{ggRandomForests} package is structured to extract intermediate data objects from \pkg{randomForestSRC} objects and generate figures using the \pkg{ggplot2} graphics package~\citep{Wickham:2009}.

Many of the figures created by the \pkg{ggRandomForests} package are also available directly from within the \pkg{randomForestSRC} package. However \pkg{ggRandomForests} offers the following advantages:

\begin{itemize}
\item Separation of data and figures: \pkg{ggRandomForests} contains functions that  operate on either the \code{randomForestSRC::rfsrc} forest object directly, or on the output from \pkg{randomForestSRC} post processing functions (i.e. \code{plot.variable}, \code{var.select}, \code{find.interaction}) to generate intermediate \pkg{ggRandomForests} data objects. functions are provide to further process these objects and plot results using the \pkg{ggplot2} graphics package. Alternatively, users can use these data objects for their own custom plotting or analysis operations.  

\item Each data object/figure is a single, self contained object. This allows simple modification and manipulation of the data or \pkg{ggplot2} objects to meet users specific needs and requirements. 

\item The use of \pkg{ggplot2} for plotting. We chose to use the \pkg{ggplot2} package for our figures to allow users flexibility in modifying the figures to their liking. Each plot function returns either a single \code{ggplot} object, or a \code{list} of \code{ggplot} objects, allowing users to use additional \pkg{ggplot2} functions or themes to modify and customize the figures to their liking.  
\end{itemize}

This document is formatted as a tutorial for using the \pkg{randomForestSRC} package for building and post-processing random forest models with the \pkg{ggRandomForests} package for investigating how the forest is constructed. In this tutorial, we use the Boston Housing Data (Section~\ref{S:data}), available in the \pkg{MASS} package~\citep{mass:2002}, to build a random forest for regression (Section~\ref{S:rfr}) and demonstrate the tools in the \pkg{ggRandomForests} package for examining the forest construction. 

Random forests are not parsimonious, but use all variables available in the construction of a response predictor. We demonstrate a random forest variable selection (Section~\ref{S:varselection}) process using the Variable Importance (Section~\ref{S:vimp}) measure (VIMP)~\citep{Breiman:2001} as well as Minimal Depth (Section~\ref{S:minimaldepth})~\citep{Ishwaran:2010}, a property derived from the construction of each tree within the forest, to assess the impact of variables on forest prediction. 

Once we have an idea of which variables we are want to investigate further, we will use variable dependence plots~\citep{Friedman:2000} to understand how a variable is related to the response (Section~\ref{S:dependence}). Marginal dependence plots (Section~\ref{S:variabledependence})  give us an idea of the overall trend of a variable/response relation, while partial dependence plots (Section~\ref{S:partialdependence}) show us a risk adjusted relation. These figures may show strongly non-linear variable/response relations that are not easily obtained through a parametric approach. We are also interested in examining variable interactions within the forest model (Section~\ref{S:interactions}). Using a minimal depth approach, we can quantify how closely variables are related within the forest, and generate marginal dependence (Section~\ref{S:coplots}) and partial dependence (Section!\ref{S:partialcoplots}) (risk adjusted) conditioning plots (coplots)~\citep{chambers:1992,cleveland:1993} to examine these interactions graphically.

\section{Data: Boston Housing Values} \label{S:data}

The Boston Housing data is a standard benchmark data set for regression models. It contains data for 506 census tracts of Boston from the 1970 census~\citep{Harrison:1978,Belsley:1980}. The data is available in multiple \proglang{R} packages, but to keep the installation dependencies for the \pkg{ggRandomForests} package down, we will use the data contained in the \pkg{MASS} package~\citep[\url{http://CRAN.R-project.org/package=MASS}]{mass:2002}, available with the base install of \proglang{R}. The following code block loads the data into the environment. We include a table of the Boston data set variable names, types and descriptions for reference when we interpret the model results.
 
\begin{Schunk}
\begin{Sinput}
R> # Load the Boston Housing data
R> data(Boston, package="MASS")
R> 
R> # Set modes correctly. For binary variables: transform to logical
R> Boston$chas <- as.logical(Boston$chas)
\end{Sinput}
\end{Schunk}

\begin{table}

\caption{\code{Boston} housing data dictionary.}
\begin{tabular}{lll}
\toprule
Variable & Description & type\\
\midrule
crim & Crime rate by town. & numeric\\
zn & Proportion of residential land zoned for lots over 25,000 sq.ft. & numeric\\
indus & Proportion of non-retail business acres per town. & numeric\\
chas & Charles River (tract bounds river). & logical\\
nox & Nitrogen oxides concentration (10 ppm). & numeric\\
\addlinespace
rm & Number of rooms per dwelling. & numeric\\
age & Proportion of units built prior to 1940. & numeric\\
dis & Distances to Boston employment center. & numeric\\
rad & Accessibility to highways. & integer\\
tax & Property-tax rate per \$10,000. & numeric\\
\addlinespace
ptratio & Pupil-teacher ratio by town. & numeric\\
black & Proportion of blacks by town. & numeric\\
lstat & Lower status of the population (percent). & numeric\\
medv & Median value of homes (\$1000s). & numeric\\
\bottomrule
\end{tabular}
\end{table}

The main objective of the Boston Housing data is to investigate variables associated with predicting the median value of homes (continuous \code{medv} response) within 506 suburban areas of Boston.

\subsection{Exploratory Data Analysis} \label{S:eda}

It is good practice to view your data before beginning an analysis, what~\cite{Tukey:1977} refers to as Exploratory Data Analysis (EDA). To facilitate this, we use \pkg{ggplot2} figures with the \code{ggplot2::facet_wrap} command to create two sets of panel plots, one for categorical variables with boxplots at each level, and one of scatter plots for continuous variables. Each variable is plotted along a selected continuous variable on the X-axis. These figures help to find outliers, missing values and other data anomalies in each variable before getting deep into the analysis. We have also created a separate \pkg{shiny} app~\citep[\url{http://shiny.rstudio.com}]{shiny:2015}, available at \url{https://ehrlinger.shinyapps.io/xportEDA}, for creating similar figures with an arbitrary data set, to make the EDA process easier for users. 

The Boston housing data consists almost entirely of continuous variables, with the exception of the ``Charles river'' logical variable. A simple EDA visualization to use for this data is a single panel plot of the continuous variables, with observation points colored by the logical variable. Missing values in our continuous variable plots are indicated by the rug marks along the x-axis, of which there are none in this data. We used the Boston housing response variable, the median value of homes (\code{medv}), for X variable.

\begin{Schunk}
\begin{Sinput}
R> # Use reshape2::melt to transform the data into long format.
R> dta <- melt(Boston, id.vars=c("medv","chas"))
R> 
R> # plot panels for each covariate colored by the logical chas variable.
R> ggplot(dta, aes(x=medv, y=value, color=chas))+
+   geom_point(alpha=.4)+
+   geom_rug(data=dta %>% filter(is.na(value)))+
+   labs(y="", x=st.labs["medv"]) +
+   scale_color_brewer(palette="Set2")+
+   facet_wrap(~variable, scales="free_y", ncol=3)
\end{Sinput}
\begin{figure}[!htb]

{\centering \includegraphics[width=\maxwidth]{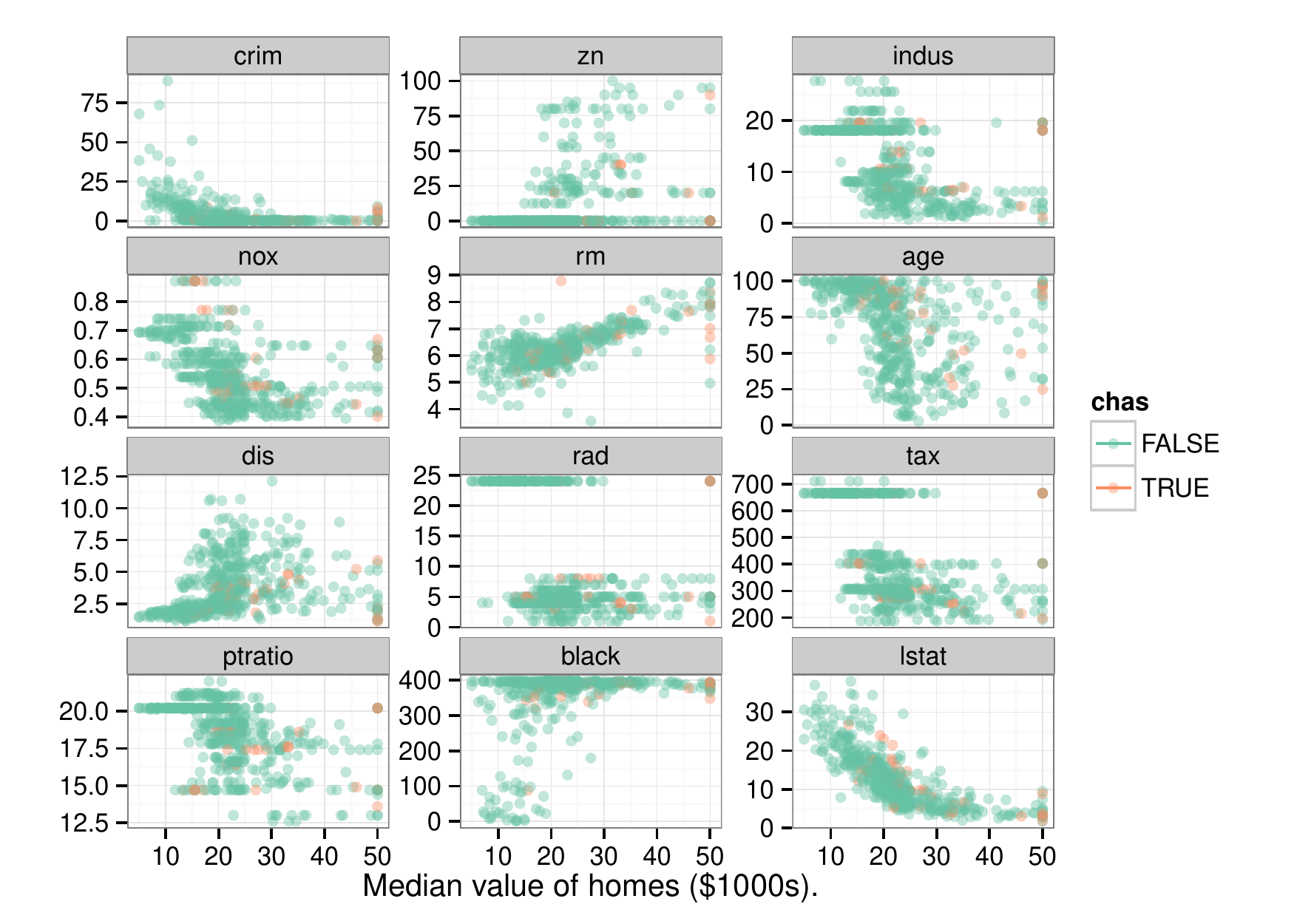} 

}

\caption[EDA variable plots]{EDA variable plots. Points indicate variable value against the median home value variable. Points are colored according to the chas variable.}\label{fig:eda}
\end{figure}
\end{Schunk}

This figure is loosely related to a pairs scatter plot~\citep{Becker:1988}, but in this case we only examine the relation between the response variable against the remainder. Plotting the data against the response also gives us a "sanity check" when viewing our model results. It's pretty obvious from this figure that we should find a strong relation between median home values and the \code{lstat} and \code{rm} variables.

\section{Random Forest - Regression} \label{S:rfr}

A Random Forest is grown by \emph{bagging}~\citep{Breiman:1996} a collection of \emph{classification and regression trees} (CART)~\citep{cart:1984}. The method uses a set of $B$ bootstrap~\citep{bootstrap:1994} samples, growing an independent tree model on each sub-sample of the population. Each tree is grown by recursively partitioning the population based on optimization of a \emph{split rule} over the $p$-dimensional covariate space. At each split, a subset of $m \le p$ candidate variables are tested for the split rule optimization, dividing each node into two daughter nodes. Each daughter node is then split again until the process reaches the \emph{stopping criteria} of either \emph{node purity} or \emph{node member size}, which defines the set of \emph{terminal (unsplit) nodes} for the tree. In regression trees, the split rule is based on minimizing the mean squared error, whereas in classification problems, the Gini index is used~\citep{Friedman:2000}.

Random Forests sort each training set observation into one unique terminal node per tree. Tree estimates for each observation are constructed at each terminal node, among the terminal node members. The Random Forest estimate for each observation is then calculated by aggregating, averaging (regression) or votes (classification), the terminal node results across the collection of $B$ trees.

For this tutorial, we grow the random forest for regression using the \code{rfsrc} command to predict the median home value (\code{medv} variable) using the remaining 13 independent predictor variables. For this example we will use the default set of $B=1000$ trees (\code{ntree} argument), $m=5$ candidate variables (\code{mtry}) for each split with a stopping criteria of at most \code{nodesize=5} observations within each terminal node. 

Because growing random forests are computationally expensive, and the \pkg{ggRandomForests} package is targeted at the visualization of random forest objects, we will use cached copies of the \pkg{randomForestSRC} objects throughout this document. We include the cached objects as data sets in the \pkg{ggRandomForests} package. The actual \code{rfsrc} calls are included in comments within code blocks. 

\begin{Schunk}
\begin{Sinput}
R> # Load the data, from the call:
R> # rfsrc_Boston <- rfsrc(medv~., data=Boston)
R> data(rfsrc_Boston)
R> 
R> # print the forest summary
R> rfsrc_Boston
\end{Sinput}
\begin{Soutput}
                         Sample size: 506
                     Number of trees: 1000
          Minimum terminal node size: 5
       Average no. of terminal nodes: 79.911
No. of variables tried at each split: 5
              Total no. of variables: 13
                            Analysis: RF-R
                              Family: regr
                      Splitting rule: regr
                % variance explained: 85.88
                          Error rate: 11.94
\end{Soutput}
\end{Schunk}

The \code{randomForestSRC::print.rfsrc} summary details the parameters used for the \code{rfsrc} call described above, and returns variance and generalization error estimate from the forest training set. The forest is built from 506 observations and 13 independent variables. It was constructed for the continuous \code{medv} variable using \code{ntree=1000} regression (\code{regr}) trees, randomly selecting 5 candidate variables at each node split, and terminating nodes with no fewer than 5 observations.

\subsection{Generalization error estimates} \label{S:error}

One advantage of Random Forests is a built in generalization error estimate. Each bootstrap sample selects approximately 63.2\% of the population on average. The remaining 36.8\% of observations, the Out-of-Bag (OOB)~\citep{BreimanOOB:1996e} sample, can be used as a hold out test set for each of the trees in the forest. An OOB prediction error estimate can be calculated for each observation by predicting the response over the set of trees which were NOT trained with that particular observation. The Out-of-Bag prediction error estimates have been shown to be nearly identical to n--fold cross validation estimates~\citep{StatisticalLearning:2009}. This feature of Random Forests allows us to obtain both model fit and validation in one pass of the algorithm.

The \code{gg_error} function operates on the \code{randomForestSRC::rfsrc} object to extract the error estimates as the forest is grown. The code block demonstrates part the \pkg{ggRandomForests} design philosophy, to create separate data objects and provide functions to operate on the data objects. The following code block first creates a \code{gg_error} object, then uses the \code{plot.gg_error} function to create a \code{ggplot} object for display.

\begin{Schunk}
\begin{Sinput}
R> # Plot the OOB errors against the growth of the forest.
R> gg_e <- gg_error(rfsrc_Boston)
R> plot(gg_e)
\end{Sinput}
\begin{figure}[!htb]

{\centering \includegraphics[width=\maxwidth]{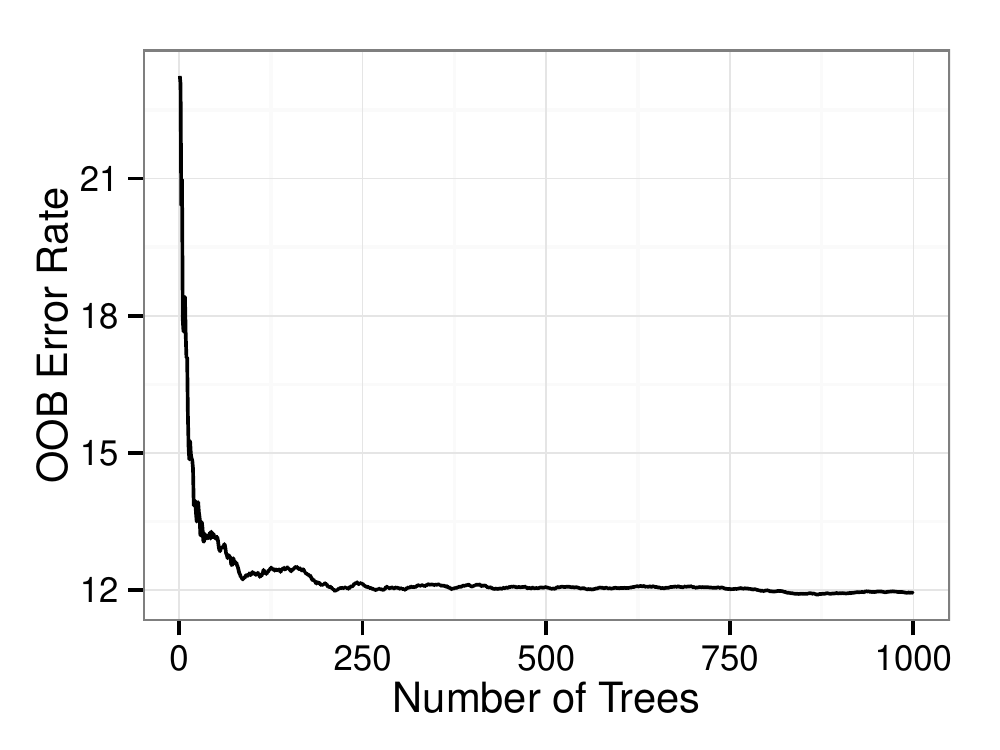} 

}

\caption[Random forest generalization error]{Random forest generalization error. OOB error convergence along the number of trees in the forest.}\label{fig:error}
\end{figure}
\end{Schunk}

This figure demonstrates that it does not take a large number of trees to stabilize the forest prediction error estimate. However, to ensure that each variable has enough of a chance to be included in the forest prediction process, we do want to create a rather large random forest of trees. 

\subsection{Random Forest Prediction} \label{S:rf-predict}

The \code{gg_rfsrc} function extracts the OOB prediction estimates from the random forest. This code block executes the the data extraction and plotting in one line, since we are not interested in holding the prediction estimates for later reuse. Also note that we add in the additional \pkg{ggplot2} command (\code{coord_cartesian}) to modify the plot object. Each of the \pkg{ggRandomForests} plot commands return \code{ggplot} objects, which we can also store for modification or reuse later in the analysis. 

\begin{Schunk}
\begin{Sinput}
R> # Plot predicted median home values.
R> plot(gg_rfsrc(rfsrc_Boston), alpha=.5)+
+   coord_cartesian(ylim=c(5,49))
\end{Sinput}
\begin{figure}[!htb]

{\centering \includegraphics[width=\maxwidth]{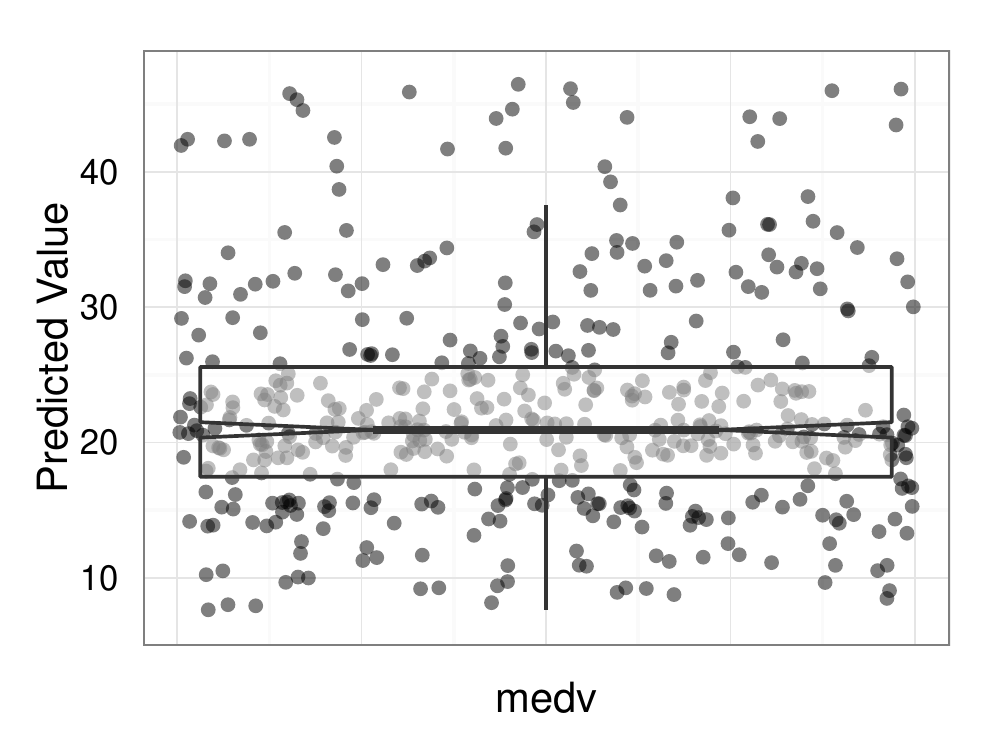} 

}

\caption[OOB predicted median home values]{OOB predicted median home values. Points are jittered to help visualize predictions for each observation. Boxplot indicates the distribution of the predicted values.}\label{fig:rfsrc}
\end{figure}
\end{Schunk}

The \code{gg_rfsrc} plot shows the predicted median home value, one point for each observation in the training set. The points are jittered around a single point on the x-axis, since we are only looking at predicted values from the forest. These estimates are Out of Bag, which are analogous to test set estimates. The boxplot is shown to give an indication of the distribution of the prediction estimates. For this analysis the figure is another model sanity check, as we are more interested in exploring the ``why'' questions for these predictions.

\section{Variable Selection} \label{S:varselection}

Random forests are not parsimonious, but use all variables available in the construction of a response predictor. Also, unlike parametric models, Random Forests do not require the explicit specification of the functional form of covariates to the response. Therefore there is no explicit p-value/significance test for variable selection with a random forest model. Instead, RF ascertain which variables contribute to the prediction through the split rule optimization, optimally choosing variables which separate observations. We use two separate approaches to explore the RF selection process, Variable Importance (Section~\ref{S:vimp}) and Minimal Depth (Section~\ref{S:minimaldepth}).

\subsection{Variable Importance.} \label{S:vimp}

\emph{Variable importance} (VIMP) was originally defined in CART using a measure involving surrogate variables (see Chapter 5 of~\cite{cart:1984}). The most popular VIMP method uses a prediction error approach involving ``noising-up'' each variable in turn. VIMP for a variable $x_v$ is the difference between prediction error when $x_v$ is noised up by randomly permuting its values, compared to prediction error under the observed values~\citep{Breiman:2001, Liaw:2002, Ishwaran:2007, Ishwaran:2008}.

Since VIMP is the difference between OOB prediction error before and after permutation, a large VIMP value indicates that misspecification detracts from the variable predictive accuracy in the forest. VIMP close to zero indicates the variable contributes nothing to predictive accuracy, and negative values indicate the predictive accuracy \emph{improves} when the variable is mispecified. In the later case, we assume noise is more informative than the true variable. As such, we ignore variables with negative and near zero values of VIMP, relying on large positive values to indicate that the predictive power of the forest is dependent on those variables. 

The \code{gg_vimp} function extracts VIMP measures for each of the variables used to grow the forest. The \code{plot.gg_vimp} function shows the variables, in VIMP rank order, from the largest (Lower Status) at the top, to smallest (Charles River) at the bottom. VIMP measures are shown using bars to compare the scale of the error increase under permutation. 

\begin{Schunk}
\begin{Sinput}
R> # Plot the VIMP rankings of independent variables.
R> plot(gg_vimp(rfsrc_Boston), lbls=st.labs)
\end{Sinput}
\begin{figure}[!htb]

{\centering \includegraphics[width=\maxwidth]{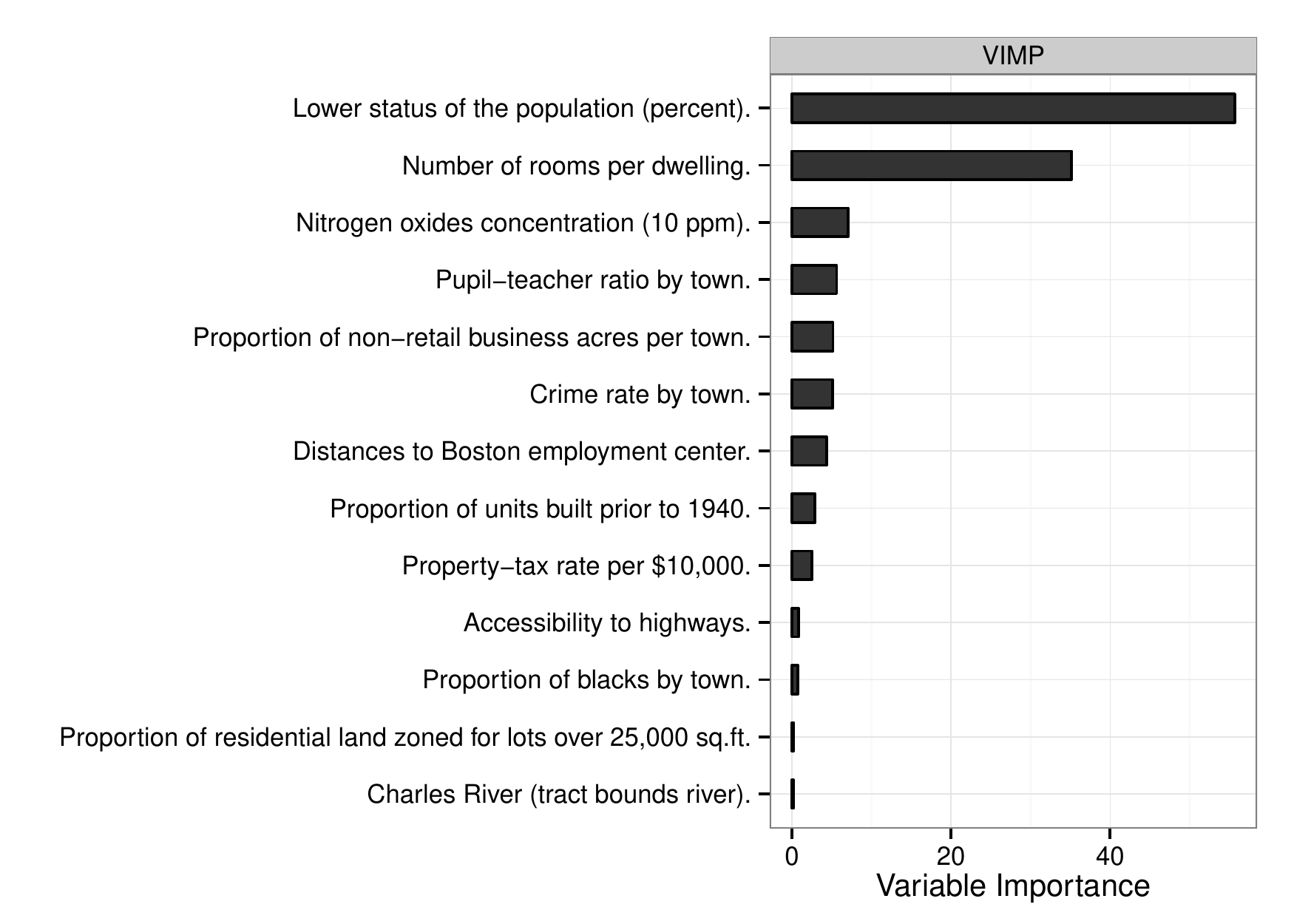} 

}

\caption[Random forest VIMP plot]{Random forest VIMP plot. Bars are colored by sign of VIMP, longer blue bars indicate more important variables.}\label{fig:vimp}
\end{figure}
\end{Schunk}

For our random forest, the top two variables (\code{lstat} and \code{rm}) have the largest VIMP, with a sizable difference to the remaining variables, which mostly have similar VIMP measure. This indicates we should focus attention on these two variables, at least, over the others.

In this example, all VIMP measures are positive, though some are small. When there are both negative and positive VIMP values, the \code{plot.gg_vimp} function will color VIMP by the sign of the measure. We use the \code{lbls} argument to pass a named \code{vector} of meaningful text descriptions to the \code{plot.gg_vimp} function, replacing the often terse variable names used by default.

\subsection{Minimal Depth.} \label{S:minimaldepth}

In VIMP, prognostic risk factors are determined by testing the forest prediction under alternative data settings, ranking the most important variables according to their impact on predictive ability of the forest. An alternative method uses inspection of the forest construction to rank variables. \emph{Minimal depth} assumes that variables with high impact on the prediction are those that most frequently split nodes nearest to the trunks of the trees (i.e. at the root node) where they partition large samples of the population. 

Within a tree, node levels are numbered based on their relative distance to the trunk of the tree (with the root at 0). Minimal depth measures the important risk factors by averaging the depth of the first split for each variable over all trees within the forest. Lower values of this measure indicate variables important in splitting large groups of patients. 

The \emph{maximal subtree} for a variable $x$ is the largest subtree whose root node splits on $x$. All parent nodes of $x$'s maximal subtree have nodes that split on variables other than $x$. The largest maximal subtree possible is at the root node. If a variable does not split the root node, it can have more than one maximal subtree, or a maximal subtree may also not exist if there are no splits on the variable. The minimal depth of a variables is a surrogate measure of predictiveness of the variable. The smaller the minimal depth, the more impact the variable has sorting observations, and therefore on the forest prediction. 

The \code{gg_minimal_depth} function is analogous to the \code{gg_vimp} function for minimal depth. Variables are ranked from most important at the top (minimal depth measure), to least at the bottom (maximal minimal depth). The vertical dashed line indicates the minimal depth threshold where smaller minimal depth values indicate higher importance and larger indicate lower importance.

The \code{randomForestSRC::var.select} call is again a computationally intensive function, as it traverses the forest finding the maximal subtree within each tree for each variable before averaging the results we use in the \code{gg_minimal_depth} call. We again use the cached object strategy here to save computational time. The \code{var.select} call is included in the comment of this code block.

\begin{Schunk}
\begin{Sinput}
R> # Load the data, from the call:
R> # varsel_Boston <- var.select(rfsrc_Boston)
R> data(varsel_Boston)
R> 
R> # Save the gg_minimal_depth object for later use.
R> gg_md <- gg_minimal_depth(varsel_Boston)
R> 
R> # plot the object
R> plot(gg_md, lbls=st.labs)
\end{Sinput}
\begin{figure}[!htb]

{\centering \includegraphics[width=\maxwidth]{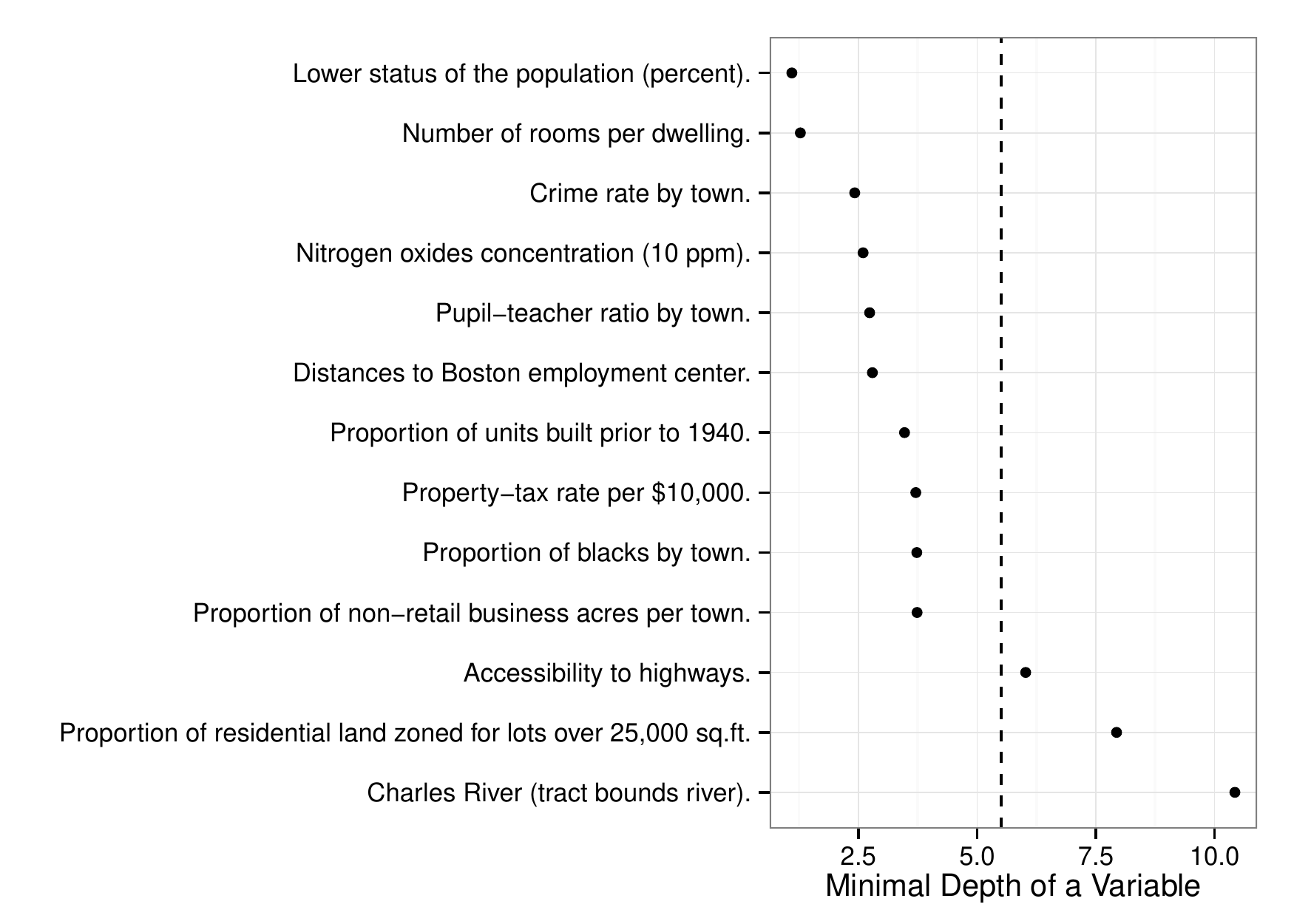} 

}

\caption[Minimal Depth variables in rank order, most important at the top]{Minimal Depth variables in rank order, most important at the top. Vertical dashed line indicates the maximal minimal depth for important variables.}\label{fig:minimaldepth}
\end{figure}
\end{Schunk}

In general, the selection of variables according to VIMP is to rather arbitrarily examine the values, looking for some point along the ranking where there is a large difference in VIMP measures. The minimal depth threshold method has a more quantitative approach to determine a selection threshold. Given minimal depth is a quantitative property of the forest construction, \cite{Ishwaran:2010} also construct an analytic threshold for evidence of variable impact. A simple optimistic threshold rule uses the mean of the minimal depth distribution, classifying variables with minimal depth lower than this threshold as important in forest prediction. The minimal depth plot for our model indicates there are ten variables which have a higher impact (minimal depth below the mean value threshold) than the remaining three. 

Since the VIMP and Minimal Depth measures use different criteria, we expect the variable ranking to be somewhat different. We use \code{gg_minimal_vimp} function to compare rankings between minimal depth and VIMP. In this call, we plot the stored \code{gg_minimal_depth} object (\code{gg_md}), which would be equivalent to calling \code{plot.gg_minimal_vimp(varsel_Boston)} or \code{plot(gg_minimal_vimp(varsel_Boston))}.

\begin{Schunk}
\begin{Sinput}
R> # gg_minimal_depth objects contain information about
R> # both minimal depth and VIMP.
R> plot.gg_minimal_vimp(gg_md)
\end{Sinput}
\begin{figure}[!htb]

{\centering \includegraphics[width=\maxwidth]{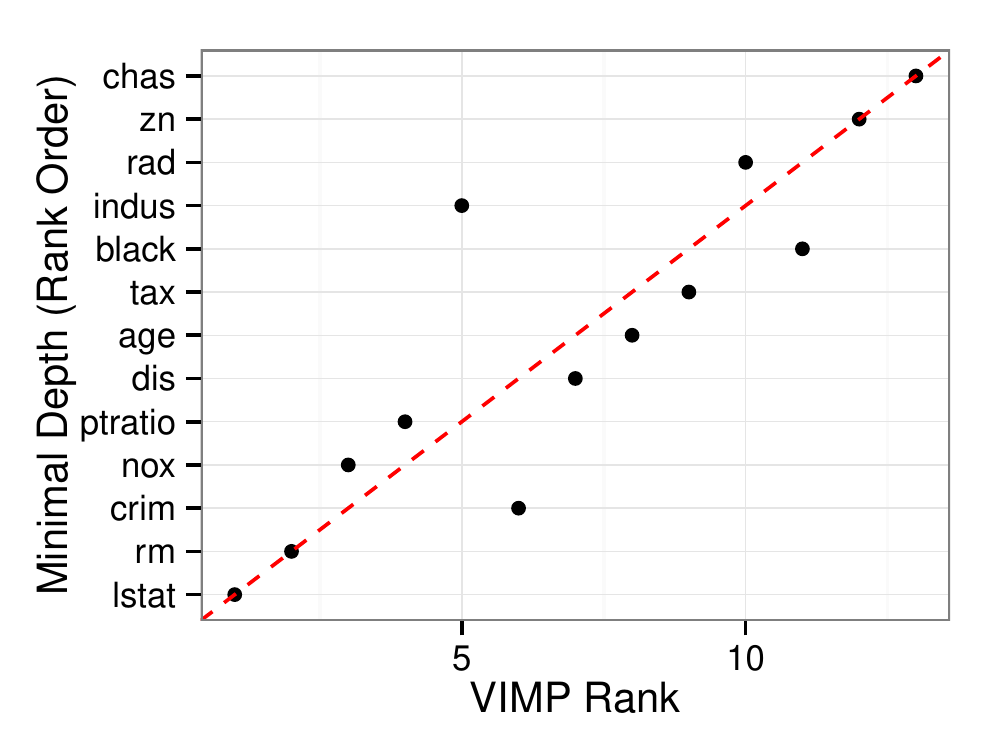} 

}

\caption[Comparing Minimal Depth and Vimp rankings]{Comparing Minimal Depth and Vimp rankings. Points on the red dashed line are ranked equivalently, points below have higher VIMP, those above have higher minimal depth ranking. Variables are colored by the sign of the VIMP measure.}\label{fig:minimalvimp}
\end{figure}
\end{Schunk}

The points along the red dashed line indicates where the measures are in agreement. Points above the red dashed line are ranked higher by VIMP than by minimal depth, indicating the variables are sensitive to misspecification. Those below the line have a higher minimal depth ranking, indicating they are better at dividing large portions of the population. The further the points are from the line, the more the discrepancy between measures. The construction of this figure is skewed towards a minimal depth approach, by ranking variables along the y-axis, though points are colored by the sign of VIMP. 

In our example, both minimal depth and VIMP indicate the strong relation of \code{lstat} and \code{rm} variables to the forest prediction, which agrees with our expectation from the EDA (Section~\ref{S:eda}) done at the beginning of this document. We now turn to investigating how these, and other variables, are related to the predicted response.

\section{Response/Variable Dependence.} \label{S:dependence}

As random forests are not a parsimonious methodology, we can use the minimal depth and VIMP measures to reduce the number of variables we need to examine to a manageable subset. We would like to know how the forest response depends on some specific variables of interest. We often choose to examine variables of interest based on the study question, or other previous knowledge. In the absence of this, we will look at variables that contribute most to the predictive accuracy of the forest.

Although often characterized as a ``black box'' method, it is possible to express a random forest in functional form. In the end the forest predictor is some function, although complex, of the predictor variables $$\hat{f}_{rf} = f(x).$$ We use graphical methods to examine the forest predicted response dependency on covariates. We again have two options, variable dependence (Section~\ref{S:variabledependence}) plots are quick and easy to generate, and partial dependence (Section~\ref{S:partialdependence}) plots are computationally intensive but give us a risk adjusted look at the dependence. 

\subsection{Variable Dependence} \label{S:variabledependence}

Variable dependence plots show the predicted response as a function of a covariate of interest, where each observation is represented by a point on the plot. Each predicted point is an individual observations, dependent on the full combination of all other covariates, not only on the covariate of interest. Interpretation of variable dependence plots can only be in general terms, as point predictions are a function of all covariates in that particular observation. However, variable dependence is straight forward to calculate, only requiring the predicted response for each observation.

We use the \code{gg_variable} function call to extract the training set variables and the predicted OOB response from \code{randomForestSRC::rfsrc} and \code{randomForestSRC::predict} objects. In the following code block, we will store the \code{gg_variable} data object for later use, as all remaining variable dependence plots can be constructed from this (\code{gg_v}) object. We will also use the minimal depth selected variables (minimal depth lower than the threshold value) from the previously stored \code{gg_minimal_depth} object (\code{gg_md$topvars}) to filter the variables of interest. 

The \code{plot.gg_variable} function call operates in the \code{gg_variable} object. We pass it the list of variables of interest (\code{xvar}) and request a single panel (\code{panel=TRUE}) to display the figures. By default, the \code{plot.gg_variable} function returns a list of \code{ggplot} objects, one figure for each variable named in \code{xvar} argument. The next three arguments are passed to internal \code{ggplot} plotting routines. The \code{se} and \code{span} arguments are used to modify the internal call to \code{ggplot2::geom_smooth} for fitting smooth lines to the data. The \code{alpha} argument lightens the coloring points in the \code{ggplot2::geom_point} call, making it easier to see point over plotting. We also demonstrate modification of the plot labels using the \code{ggplot2::labs} function.

\begin{Schunk}
\begin{Sinput}
R> # Create the variable dependence object from the random forest
R> gg_v <- gg_variable(rfsrc_Boston)
R> 
R> # We want the top ranked minimal depth variables only,
R> # plotted in minimal depth rank order. 
R> xvar <- gg_md$topvars
R> 
R> # plot the variable list in a single panel plot
R> plot(gg_v, xvar=xvar, panel=TRUE, 
+      se=.95, span=1.2, alpha=.4)+
+   labs(y=st.labs["medv"], x="")
\end{Sinput}
\begin{figure}[!htb]

{\centering \includegraphics[width=\maxwidth]{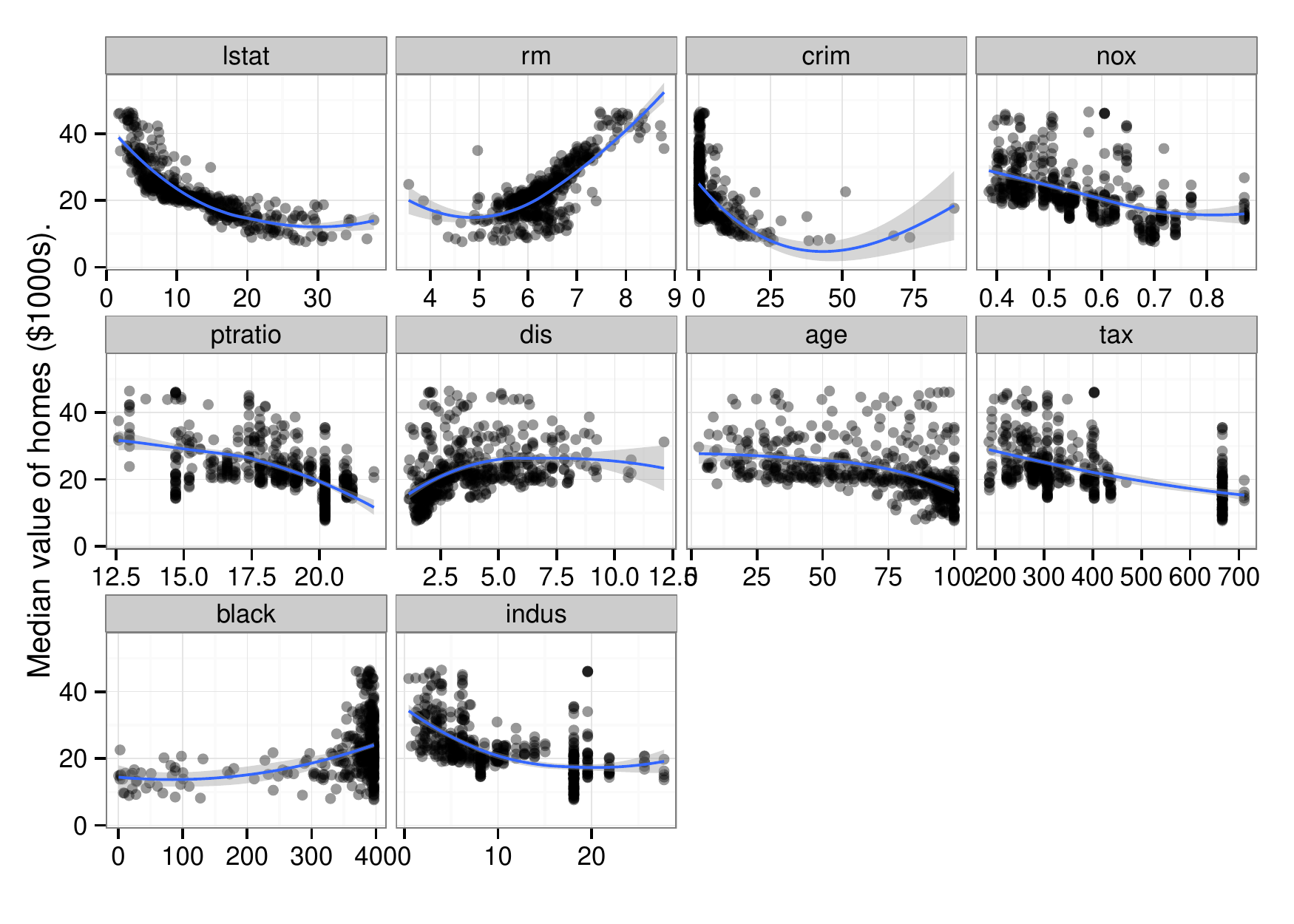} 

}

\caption[Variable dependence plot]{Variable dependence plot. Individual case predictions are marked with points. Loess smooth curve indicates the trend as the variables increase with shaded 95\% confidence band.}\label{fig:variable}
\end{figure}
\end{Schunk}

This figure looks very similar to the EDA (Section~\ref{S:eda}) figure, although with transposed axis as we plot the response variable on the y-axis. The closer the panels match, the better the RF prediction. The panels are sorted to match the order of variables in the \code{xvar} argument and include a smooth loess line~\citep{cleveland:1981, cleveland:1988}, with 95\% shaded confidence band, to indicates the trend of the prediction dependence over the covariate values.

There is not a convenient method to panel scatter plots and boxplots together, so we recommend creating panel plots for each variable type separately. The Boston housing data does contain a single categorical variable, the Charles river logical variable. Variable dependence plots for categorical variables are constructed using boxplots to show the distribution of the predictions within each category. Although the Charles river variable has the lowest importance scores in both VIMP and minimal depth measures, we include the variable dependence plot as an example of categorical variable dependence.

\begin{Schunk}
\begin{Sinput}
R> plot(gg_v, xvar="chas", points=FALSE,
+      se=FALSE, notch=TRUE, alpha=.4)+
+   labs(y=st.labs["medv"])
\end{Sinput}
\begin{figure}[!htb]

{\centering \includegraphics[width=\maxwidth]{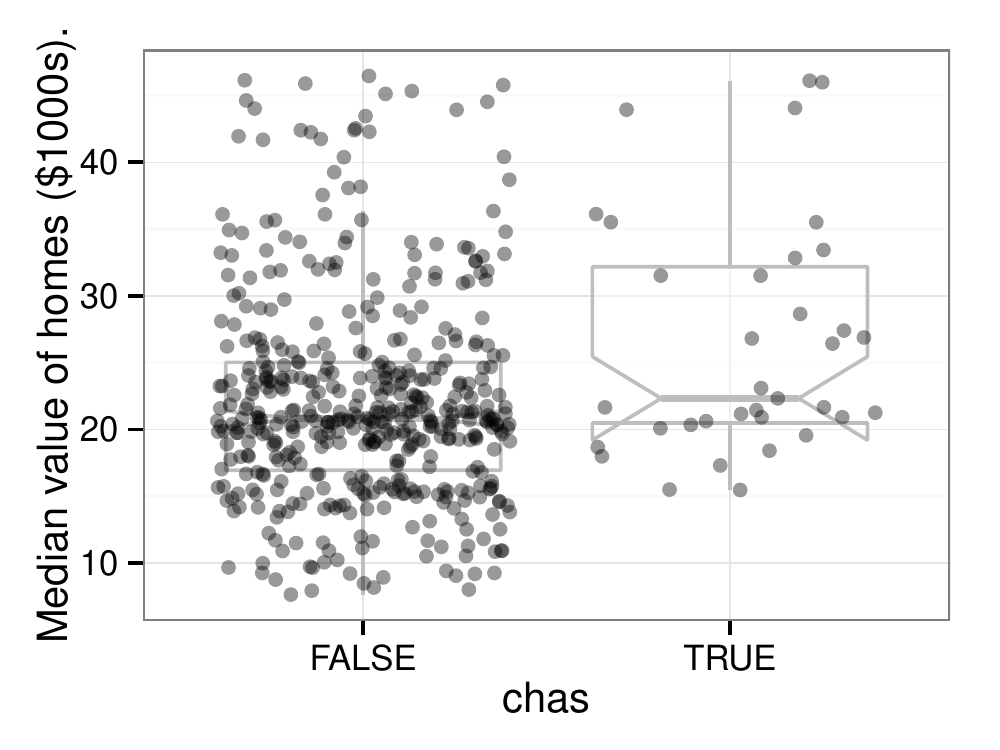} 

}

\caption[Variable dependence for Charles River logical variable]{Variable dependence for Charles River logical variable.}\label{fig:chas}
\end{figure}
\end{Schunk}

The figure shows that most housing tracts do not border the Charles river (\code{chas=FALSE}), and comparing the distributions of the predicted median housing values indicates no significant difference in home values. This reinforces the findings in both VIMP and Minimal depth, the Charles river variable has very little impact on the forest prediction of median home values.  

\subsection{Partial Dependence.} \label{S:partialdependence}

Partial variable dependence plots are a risk adjusted alternative to variable dependence. Partial plots are generated by integrating out the effects of all variables beside the covariate of interest. Partial dependence data are constructed by selecting points evenly spaced along the distribution of the $X$ variable of interest. For each value ($X = x$), we calculate the average RF prediction over all other covariates in $X$ by
$$ \tilde{f}(x) = \frac{1}{n} \sum_{i = 1}^n \hat{f}(x, x_{i, o}), $$
where $\hat{f}$ is the predicted response from the random forest and $x_{i, o}$ is the value for all other covariates other than $X = x$ for the observation $i$~\citep{Friedman:2000}. Essentially, we average a set of predictions for each observation in the training set at the value of $X=x$. We repeating the process for a sequence of $X=x$ values to generate the estimated points to create a partial dependence plot. 

Partial plots are another computationally intensive analysis, especially when there are a large number of observations. We again turn to our data caching strategy here. The default parameters for the \code{randomForestSRC::plot.variable} function generate partial dependence estimates at \code{npts=25} points (the default value) along the variable of interest. For each point of interest, the \code{plot.variable} function averages \code{n} response predictions. This is repeated for each of the variables of interest and the results are returned for later analysis. 

\begin{Schunk}
\begin{Sinput}
R> # Load the data, from the call:
R> # partial_Boston <- plot.variable(rfsrc_Boston, 
R> #                                 xvar=gg_md$topvars, 
R> #                                 partial=TRUE, sorted=FALSE, 
R> #                                 show.plots = FALSE )
R> data(partial_Boston)
R> 
R> # generate a list of gg_partial objects, one per xvar.
R> gg_p <- gg_partial(partial_Boston)
R> 
R> # plot the variable list in a single panel plot
R> plot(gg_p, xvar=xvar, panel=TRUE, se=FALSE) +
+   labs(y=st.labs["medv"], x="")
\end{Sinput}
\begin{figure}[!htb]

{\centering \includegraphics[width=\maxwidth]{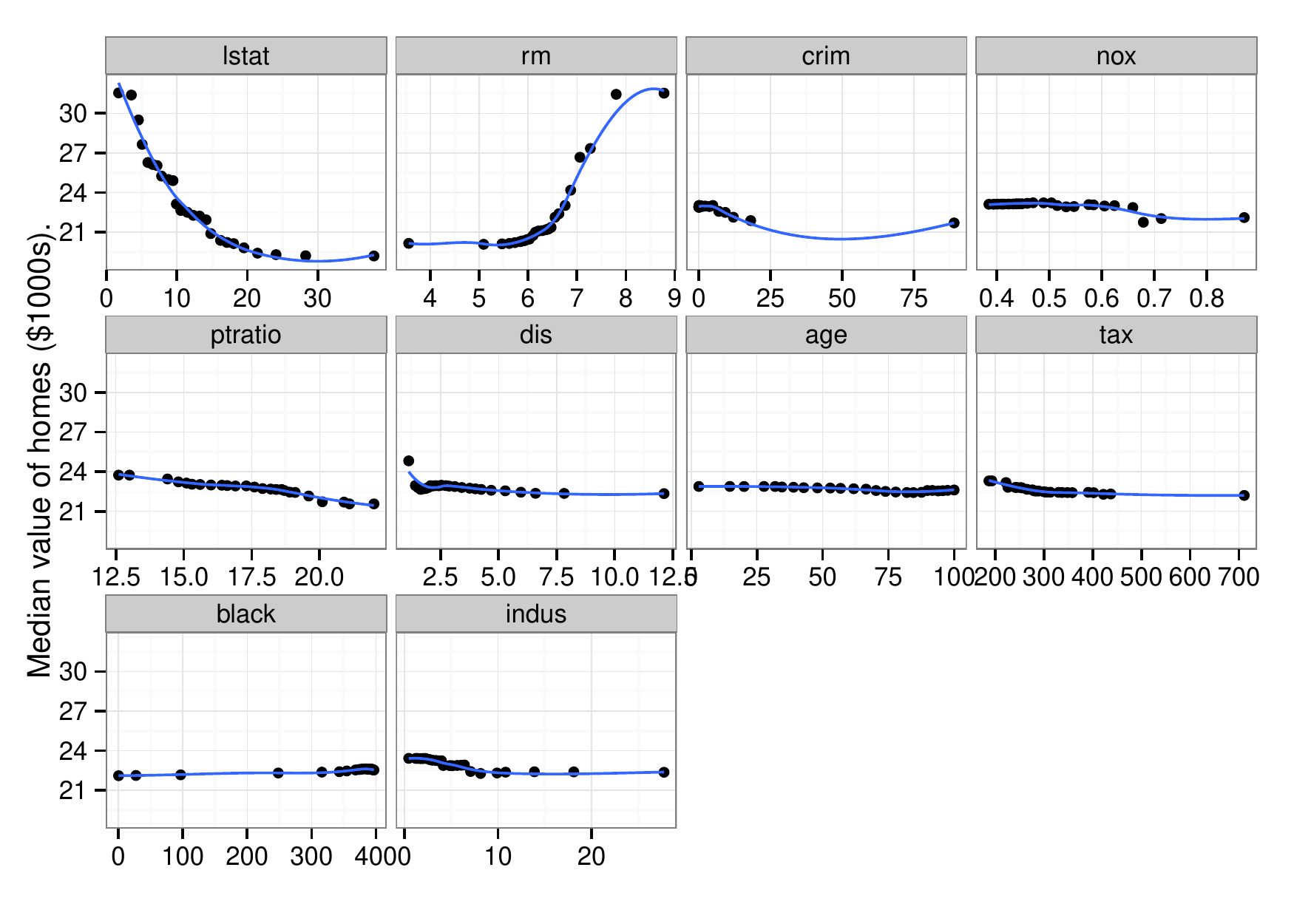} 

}

\caption[Partial dependence panels]{Partial dependence panels. Risk adjusted variable dependence for variables in minimal depth rank order.}\label{fig:partial}
\end{figure}
\end{Schunk}

We again order the panels by minimal depth ranking. We see again how the \code{lstat} and \code{rm} variables are strongly related to the median value response, making the partial dependence of the remaining variables look flat. We also see strong nonlinearity of these two variables. The \code{lstat} variable looks rather quadratic, while the \code{rm} shape is more complex.  

We could stop here, indicating that the RF analysis has found these ten variables to be important in predicting the median home values. That increasing \code{lstat} (percentage population of lower status) values are associated with decreasing median home values (\code{medv}) and increasing `rm > 6` (number of rooms $> 6$) are associated with increasing median home values. However, we may also be interested in investigating how these variables work together to help improve the random forest prediction of median home values.

\section{Variable Interactions} \label{S:interactions}

Using the different variable dependence measures, it is also possible to calculate measures of pairwise interactions among variables. Recall that minimal depth measure is defined by averaging the tree depth of variable $i$ relative to the root node. To detect interactions, this calculation can be modified to measure the minimal depth of a variable $j$ with respect to the maximal subtree for variable $i$~\citep{Ishwaran:2010, Ishwaran:2011}.

The \code{randomForestSRC::find.interaction} function traverses the forest, calculating all pairwise minimal depth interactions, and returns a $p \times p$ matrix of interaction measures. For each row, the diagonal terms are are related to the minimal depth relative to the root node, though normalized to minimize scaling issues. Each off diagonal minimal depth term is relative to the diagonal term on that row. Small values indicate that an off diagonal term typically splits close to the diagonal term, indicating an forest split proximity of the two variables.

The \code{gg_interaction} function wraps the \code{find.interaction} matrix for use with the provided plot and print functions. The \code{xvar} argument indicates which variables we're interested in looking at. We again use the cache strategy, and collect the figures together using the \code{panel=TRUE} option.

\begin{Schunk}
\begin{Sinput}
R> # Load the data, from the call:
R> # interaction_Boston <- find.interactions(rfsrc_Boston)
R> data(interaction_Boston)
R> 
R> # Plot the results in a single panel.
R> plot(gg_interaction(interaction_Boston), 
+      xvar=gg_md$topvars, panel=TRUE)
\end{Sinput}
\begin{figure}[!htb]

{\centering \includegraphics[width=\maxwidth]{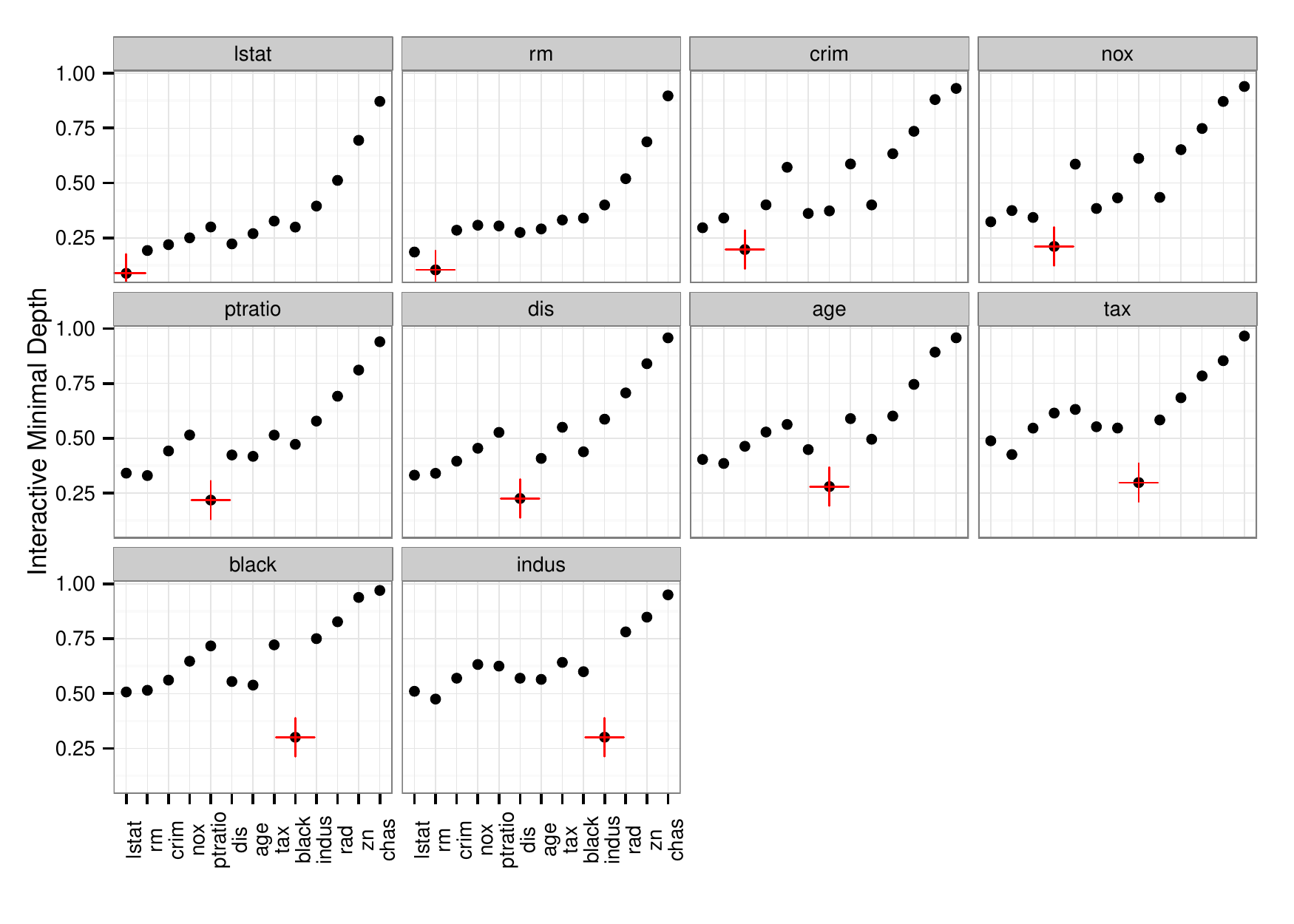} 

}

\caption[Minimal depth variable interactions]{Minimal depth variable interactions. Reference variables are marked with red cross in each panel. Higher values indicate lower interactivity with reference variable.}\label{fig:interactions}
\end{figure}
\end{Schunk}

The \code{gg_interaction} figure plots the interactions for the target variable (shown in the red cross) with interaction scores for all remaining variables. We expect the covariate with lowest minimal depth (\code{lstat}) to be associated with almost all other variables, as it typically splits close to the root node, so viewed alone it may not be as informative as looking at a collection of interactive depth plots. Scanning across the panels, we see each successive target depth increasing, as expected. We also see the interactive variables increasing with increasing target depth. Of interest here is the interaction of \code{lstat} with the \code{rm} variable shown in the \code{rm} panel. Aside from these being the strongest variables by both measures, this interactive measure indicates the strongest connection between variables.

\section{Coplots} \label{S:coplots}
Conditioning plots (coplots)~\citep{chambers:1992, cleveland:1993}  are a powerful visualization tool to efficiently study how a response depends on two or more variables~\citep{cleveland:1993}. The method allows us to view data by grouping observations on some conditional membership. The simplest example involves a categorical variable, where we plot our data conditional on class membership, for instance on the Charles river logical variable. We can view a coplot as a stratified variable dependence plot, indicating trends in the RF prediction results within panels of group membership.

Conditional membership with a continuous variable requires stratification at some level. Often we can make these stratification along some feature of the variable, for instance a variable with integer values, or 5 or 10 year age group cohorts. However in the variables of interest in our Boston housing example, we have no "logical" stratification indications. Therefore we will arbitrarily stratify our variables into 6 groups of roughly equal population size using the \code{quantile_pts} function. We pass the break points located by \code{quantile_pts} to the \code{cut} function to create grouping intervals, which we can then add to the \code{gg_variable} object before plotting with the \code{plot.gg_variable} function. The simple modification to convert variable dependence plots into condition variable dependence plots is to use the \code{ggplot2::facet_wrap} command to generate a panel for each grouping interval.

We start by examining the predicted median home value as a function of \code{lstat} conditional on membership within 6 groups of \code{rm} ``intervals''. 
\begin{Schunk}
\begin{Sinput}
R> # Find the rm variable points to create 6 intervals of roughly 
R> # equal size population
R> rm_pts <- quantile_pts(rfsrc_Boston$xvar$rm, groups=6, intervals=TRUE)
R> 
R> # Pass these variable points to create the 6 (factor) intervals
R> rm_grp <- cut(rfsrc_Boston$xvar$rm, breaks=rm_pts)
R> 
R> # Append the group factor to the gg_variable object
R> gg_v$rm_grp <- rm_grp
R> 
R> # Modify the labels for descriptive panel titles 
R> levels(gg_v$rm_grp) <- paste("rm in ", levels(gg_v$rm_grp), sep="")
R> 
R> # Create a variable dependence (co)plot, faceted on group membership.
R> plot(gg_v, xvar = "lstat", smooth = TRUE, 
+      method = "loess", span=1.5, alpha = .5, se = FALSE) + 
+   labs(y = st.labs["medv"], x=st.labs["lstat"]) + 
+   theme(legend.position = "none") + 
+   scale_color_brewer(palette = "Set3") + 
+   facet_wrap(~rm_grp)
\end{Sinput}
\begin{figure}[!htb]

{\centering \includegraphics[width=\maxwidth]{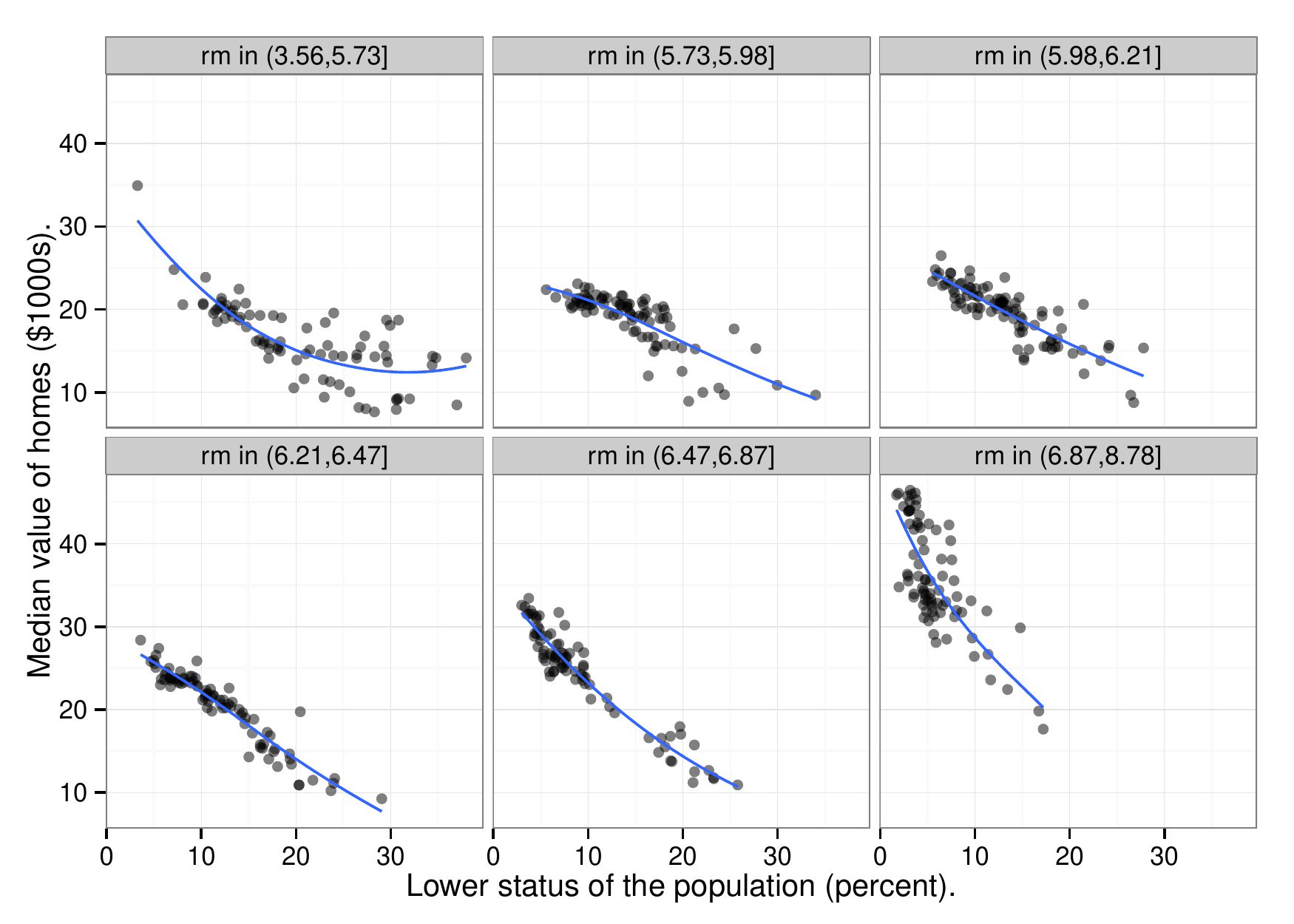} 

}

\caption[Variable Coplots]{Variable Coplots. Predicted median home values as a function of percentage of lower status population, stratified by average number of rooms groups.}\label{fig:coplots}
\end{figure}
\end{Schunk}

Each point in this figure is the predicted median home value response plotted against \code{lstat} value conditional on \code{rm} being on the interval specified. We again use the smooth loess curve to get an idea of the trend within each group. Overall, median values continue to decrease with increasing \code{lstat}, and increases with increasing \code{rm}. In addition to trends, we can also examine the conditional distribution of variables. Note that smaller homes (\code{rm}) in high status (lower \code{lstat}) neighborhoods still have high predicted median values, and that there are more large homes in the higher status neighborhoods (bottom right panel).

A single coplot gives us a grouped view of a variable (\code{rm}), along the primary variable dimension (\code{lstat}). To get a better feel for how the response depends on both variables, it is instructive to look at the complement coplot. We repeat the previous coplot process, predicted median home value as a function of the \code{rm} variable, conditional on membership within 6 groups \code{lstat} intervals. 

\begin{Schunk}
\begin{Sinput}
R> # Find the lstat variable points to create 6 intervals of roughly 
R> # equal size population
R> lstat_pts <- quantile_pts(rfsrc_Boston$xvar$lstat, groups=6, intervals=TRUE)
R> 
R> # Pass these variable points to create the 6 (factor) intervals
R> lstat_grp <- cut(rfsrc_Boston$xvar$lstat, breaks=lstat_pts)
R> 
R> # Append the group factor to the gg_variable object
R> gg_v$lstat_grp <- lstat_grp
R> 
R> # Modify the labels for descriptive panel titles 
R> levels(gg_v$lstat_grp) <- paste("lstat in ", levels(gg_v$lstat_grp), " (%)",sep="")
R> 
R> # Create a variable dependence (co)plot, faceted on group membership.
R> plot(gg_v, xvar = "rm", smooth = TRUE, 
+      method = "loess", span=1.5, alpha = .5, se = FALSE) + 
+   labs(y = st.labs["medv"], x=st.labs["rm"]) + 
+   theme(legend.position = "none") + 
+   scale_color_brewer(palette = "Set3") + 
+   #scale_shape_manual(values = event.marks, labels = event.labels)+ 
+   facet_wrap(~lstat_grp)
\end{Sinput}
\begin{figure}[!htb]

{\centering \includegraphics[width=\maxwidth]{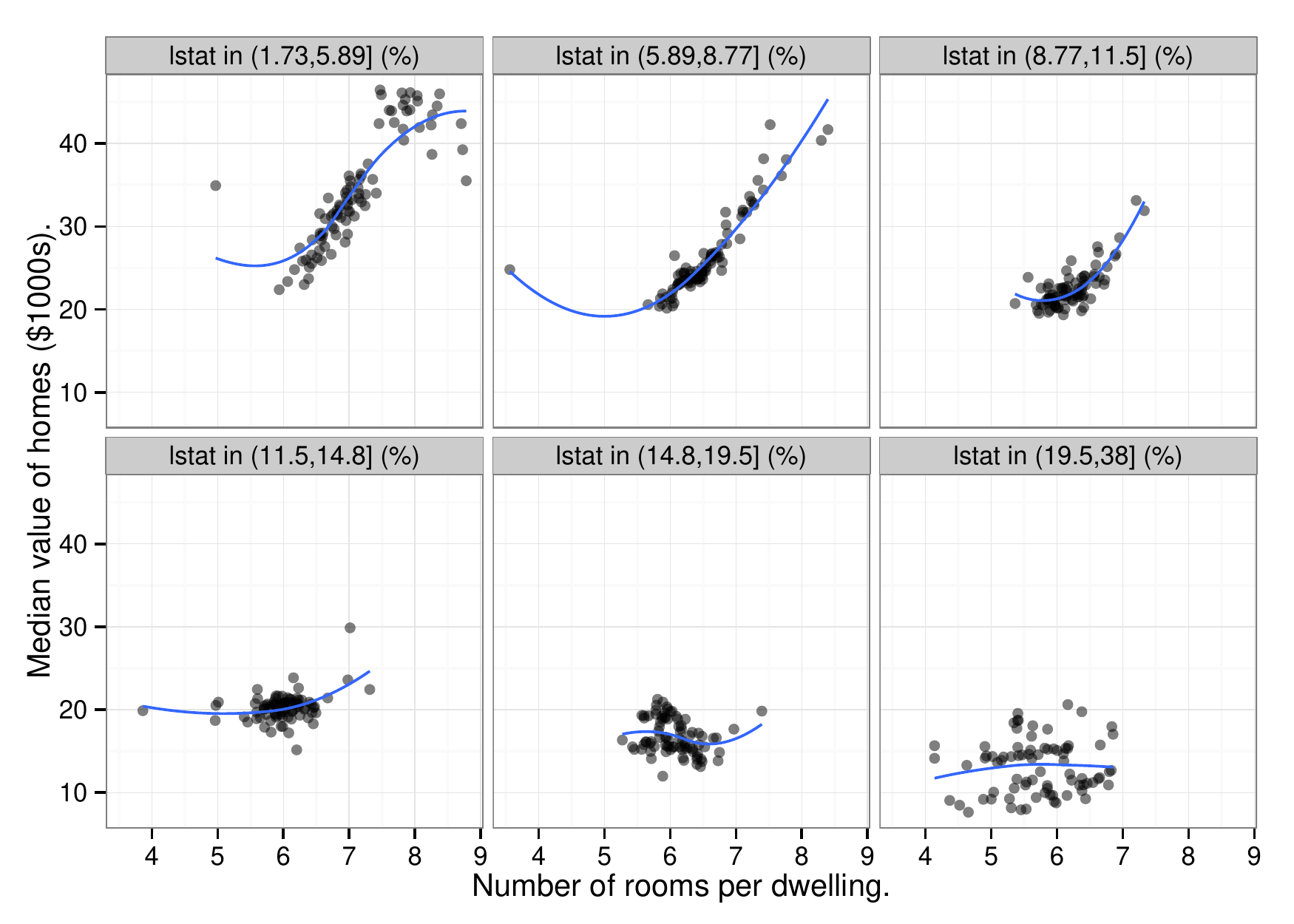} 

}

\caption[Variable Coplots]{Variable Coplots. Predicted median home value as a function of average number of rooms, stratified by percentage of lower status groups.}\label{fig:coplots2}
\end{figure}
\end{Schunk}

We get similar information from this view, predicted median home values decrease with increasing \code{lstat} percentage and decreasing \code{rm}. However viewed together we get a better sense of how the \code{lstat} and \code{rm} variables work together (interact) in the median value prediction.

Note that typically \cite{cleveland:1993} conditional plots for continuous variables included overlapping intervals along the grouped variable. We chose to use mutually exclusive continuous variable intervals for multiple reasons:
 
 \begin{itemize}
 \item Simplicity - We can create the coplot figures directly from the \code{gg_variable} object by adding a conditional group column directly to the object.

 \item Interpretability - We find it easier to interpret and compare the panels if each observation is only in a single panel.

\item Clarity - We prefer using more space for the data portion of the figures than typically displayed in the \code{coplot} function available in base R, which require the bar plot to present the overlapping segments.
\end{itemize}

It is still possible to augment the \code{gg_variable} to include overlapping conditional membership with continuous variables by duplicating rows of the object, and setting the correct conditional group membership. The \code{plot.gg_variable} function recipe above could then be used to generate the panel plot, with panels ordered according to the factor levels of the grouping variable. We leave this as an exercise for the reader.

\subsection{Partial dependence coplots} \label{S:partialcoplots}

By characterizing conditional plots as stratified variable dependence plots, the next logical step would be to generate an analogous conditional partial dependence plot. The process is similar to variable dependence coplots, first determine conditional group membership, then calculate the partial dependence estimates on each subgroup using the \code{randomForestSRC::plot.variable} function with a the \code{subset} argument for each grouped interval. The \code{ggRandomForests::gg_partial_coplot} function is a wrapper for generating a conditional partial dependence data object. Given a random forest (\code{randomForestSRC::rfsrc} object) and a \code{groups} vector for conditioning the training data set observations, \code{gg_partial_coplot} calls the \code{randomForestSRC::plot.variable} function for a set of training set observations conditional on \code{groups} membership. The function returns a \code{gg_partial_coplot} object, a sub class of the \code{gg_partial} object, which can be plotted with the \code{plot.gg_partial} function.

The following code block will generate the data object for creating partial dependence coplot of the predicted median home value as a function of \code{lstat} conditional on membership within the 6 groups of \code{rm} "intervals" that we examined in the previous section.

\begin{Schunk}
\begin{Sinput}
R> partial_coplot_Boston <- gg_partial_coplot(rfsrc_Boston, xvar="lstat", 
+                                            groups=rm_grp,
+                                            show.plots=FALSE)
\end{Sinput}
\end{Schunk}

Since the \code{gg_partial_coplot} makes a call to \code{randomForestSRC::plot.variable} for each group (6) in the conditioning set, we again resort to the data caching strategy, and load the stored result data from the \code{ggRandomForests} package. We modify the legend label to indicate we're working with groups of the  ``Room'' variable, and use the \code{palette="Set1"} from the \pkg{RColorBrewer} package~\citep{rcolorbrewer:2014}  to choose a nice color theme for displaying the six curves.

\begin{Schunk}
\begin{Sinput}
R> # Load the stored partial coplot data.
R> data(partial_coplot_Boston)
R> 
R> # Partial coplot
R> plot(partial_coplot_Boston, se=FALSE)+
+   labs(x=st.labs["lstat"], y=st.labs["medv"], 
+        color="Room", shape="Room")+
+   scale_color_brewer(palette="Set1")
\end{Sinput}
\begin{figure}[!htb]

{\centering \includegraphics[width=\maxwidth]{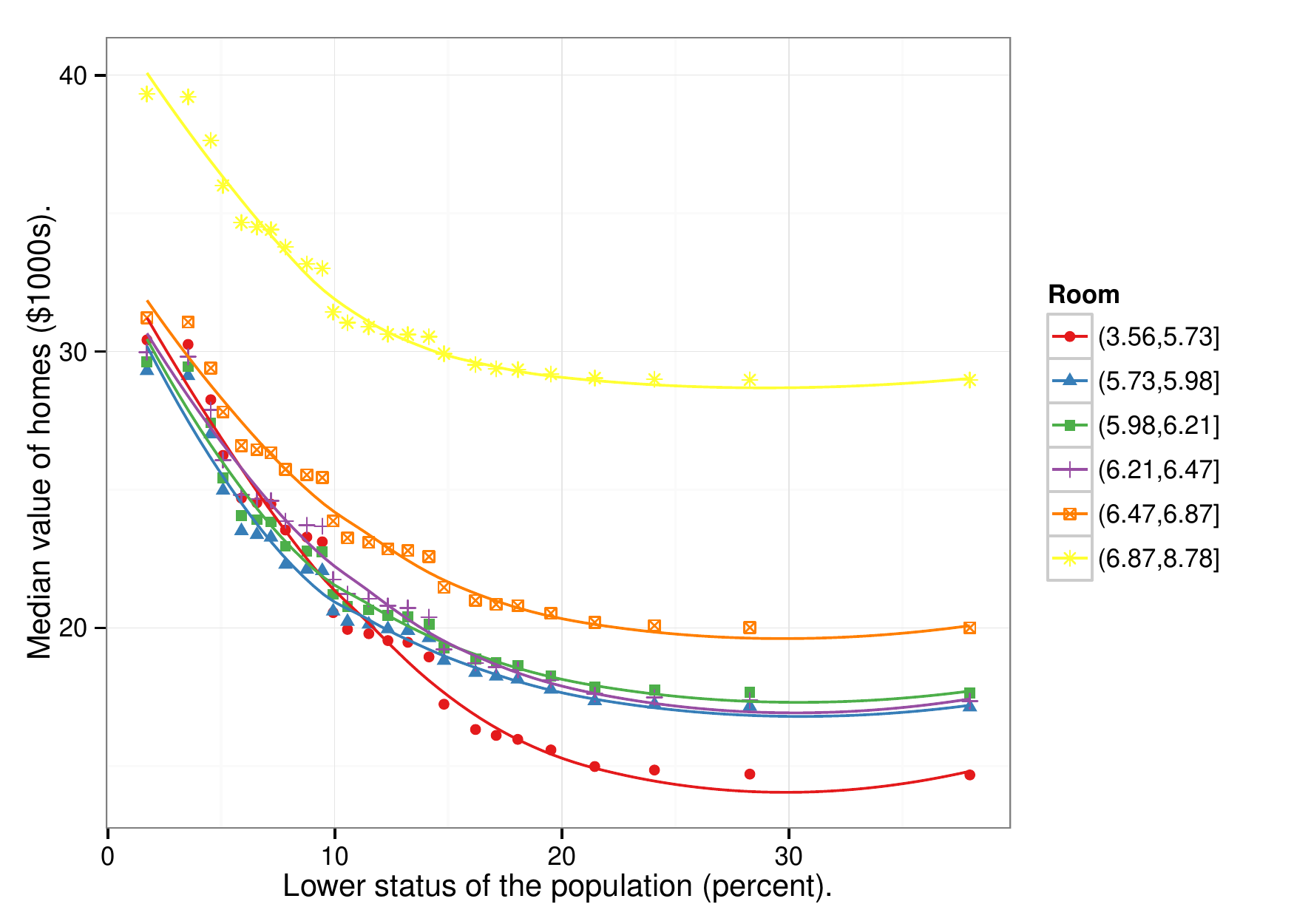} 

}

\caption[Partial Coplots]{Partial Coplots. Risk adjusted predicted median value as a function of Lower Status, conditional on groups of average number of rooms.}\label{fig:prtl-coplots}
\end{figure}
\end{Schunk}

Unlike variable dependence coplots, we do not need to use a panel format for partial dependence coplots because we are looking risk adjusted estimates (points) instead of population estimates. The figure has a loess curve through the point estimates conditional on the \code{rm} interval groupings. The figure again indicates that larger homes (\code{rm} from 6.87 and up, shown in yellow) have a higher median value then the others. In neighborhoods with higher \code{lstat} percentage, the Median values decrease with \code{rm} until it stabilizes from the intervals between 5.73 and 6.47, then decreases again for values smaller than 5.73. In lower \code{lstat} neighborhoods, the effect of smaller \code{rm} is not as noticeable.

We can view the partial coplot curves as slices along a surface viewed into the page, either along increasing or decreasing \code{rm} values. This is made more difficult by our choice to select groups of similar population size, as the curves are not evenly spaced along the \code{rm} variable. We return to this problem in the next section. 

We also construct the complement view, for partial dependence coplot of the predicted median home value as a function of \code{rm} conditional on membership within the 6 groups of \code{lstat}  ``intervals'', and cache the following \code{gg_partial_coplot} data call, and plot the results with the \code{plot.gg_variable} call:

\begin{Schunk}
\begin{Sinput}
R> partial_coplot_Boston2 <- gg_partial_coplot(rfsrc_Boston, xvar="rm", 
+                                             groups=lstat_grp,
+                                             show.plots=FALSE)
\end{Sinput}
\end{Schunk}

\begin{Schunk}
\begin{Sinput}
R> # Load the stored partial coplot data.
R> data(partial_coplot_Boston2)
R> 
R> # Partial coplot
R> plot(partial_coplot_Boston2, se=FALSE)+
+   labs(x=st.labs["rm"], y=st.labs["medv"], 
+        color="Lower Status", shape="Lower Status")+
+   scale_color_brewer(palette="Set1")
\end{Sinput}
\begin{figure}[!htb]

{\centering \includegraphics[width=\maxwidth]{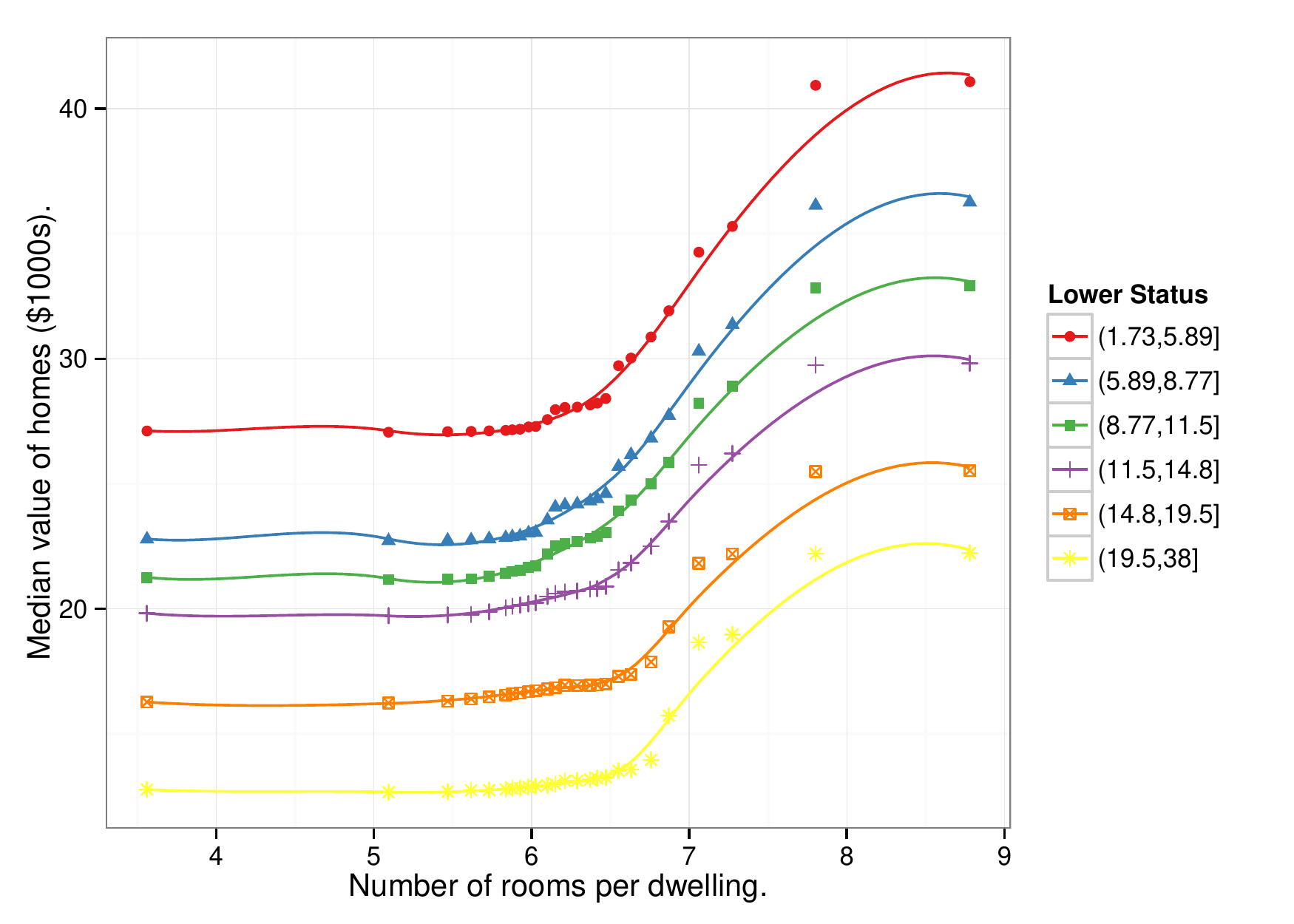} 

}

\caption[Partial Coplots]{Partial Coplots. Risk adjusted predicted median value as a function of average number of rooms, conditional on groups of percentage of lower status population.}\label{fig:prtl-coplots2}
\end{figure}
\end{Schunk}

This figure indicates that the median home value does not change much until the \code{rm} increases above 6.5, then flattens again above 8, regardless of the \code{lstat} value. This agrees well with the \code{rm} partial plot(Section~\ref{S:partialdependence}) shown earlier. Again, care must be taken in interpreting the even spacing of these curves along the percentage of \code{lstat} groupings, as again, we chose these groups to have similar sized populations, not to be evenly spaced along the \code{lstat} variable. 

\section{Partial plot surfaces} \label{S:plotsurface}

Visualizing two dimensional projections of three dimensional data is difficult, though there are tools available to make the data more understandable. To make the interplay of lower status and average room size a bit more understandable, we will generate a contour partial plot of the median home values. We could generate this figure with the coplot data we already have, but the resolution would be a bit strange. To generate the plot of \code{lstat} conditional on \code{rm} groupings, we would end up with contours over a grid of \code{lstat}= $25 \times$ \code{rm}= $6$, for the alternative \code{rm} conditional on \code{lstat} groups, we'd have the transpose grid of \code{lstat}= $6 \times$  \code{rm}= $25$. 

Since we are already using the data caching strategy, we will generate another set of \code{gg_partial} objects with increased resolution in both the \code{lstat} and \code{rm} dimensions. For this exercise, we will find 50 points evenly spaced along the \code{rm} variable values, and generate a partial plot curve for each point. For these partial plots, we will evaluate the risk adjusted median home value over `npts=50` points along the \code{lstat} variable. This code block finds 50 \code{rm} values evenly spaced along the distribution of \code{rm}.
\begin{Schunk}
\begin{Sinput}
R> # Find the quantile points to create 50 cut points
R> rm_pts <- quantile_pts(rfsrc_Boston$xvar$rm, groups=50)
\end{Sinput}
\end{Schunk}

We use the following data call to generate the partial plots with the \code{randomForestSRC::plot.variable} call. Within the lapply call, we use scope to modify the value of the \code{rm} variable within the \code{rfsrc_Boston} training set. Since all values in the training set are the same, the averaged value of \code{rm} places each partial plot curve at a specific value of \code{rm}. This code block took about 20 minutes to run on a quad core Mac Air using a single processor.  

The cached data is stored in the \code{partial_Boston_surf} data set in the \code{ggRandomForests} package. The data set is a \code{list} of 50 \code{plot.variable} objects. This code block loads the data, converts the \code{plot.variable} objects to \code{gg_partial} objects, attaches numeric values for the \code{rm} variable, and generates the contour plot.

\begin{Schunk}
\begin{Sinput}
R> # Generate the gg_partial_coplot data object
R> system.time(partial_Boston_surf <- lapply(rm_pts, function(ct){
+   rfsrc_Boston$xvar$rm <- ct
+   plot.variable(rfsrc_Boston, xvar = "lstat", time = 1,
+                 npts = 50, show.plots = FALSE, 
+                 partial = TRUE)
+ }))
R> #     user   system  elapsed 
R> # 1109.641   76.516 1199.732 
\end{Sinput}
\end{Schunk}

\begin{Schunk}
\begin{Sinput}
R> # Load the stored partial coplot data.
R> data(partial_Boston_surf)
R> 
R> # Instead of groups, we want the raw rm point values,
R> # To make the dimensions match, we need to repeat the values
R> # for each of the 50 points in the lstat direction
R> rm.tmp <- do.call(c,lapply(rm_pts, 
+                            function(grp){rep(grp, 50)}))
R> 
R> # Convert the list of plot.variable output to 
R> partial_surf <- do.call(rbind,lapply(partial_Boston_surf, gg_partial))
R> 
R> # attach the data to the gg_partial_coplot
R> partial_surf$rm <- rm.tmp
R> 
R> # ggplot2 contour plot of x, y and z data.
R> ggplot(partial_surf, aes(x=lstat, y=rm, z=yhat))+
+   stat_contour(aes(colour = ..level..), binwidth = .5)+
+   labs(x=st.labs["lstat"], y=st.labs["rm"], 
+        color="Median Home Values")+
+   scale_colour_gradientn(colours=topo.colors(10))
\end{Sinput}
\begin{figure}[!htb]

{\centering \includegraphics[width=\maxwidth]{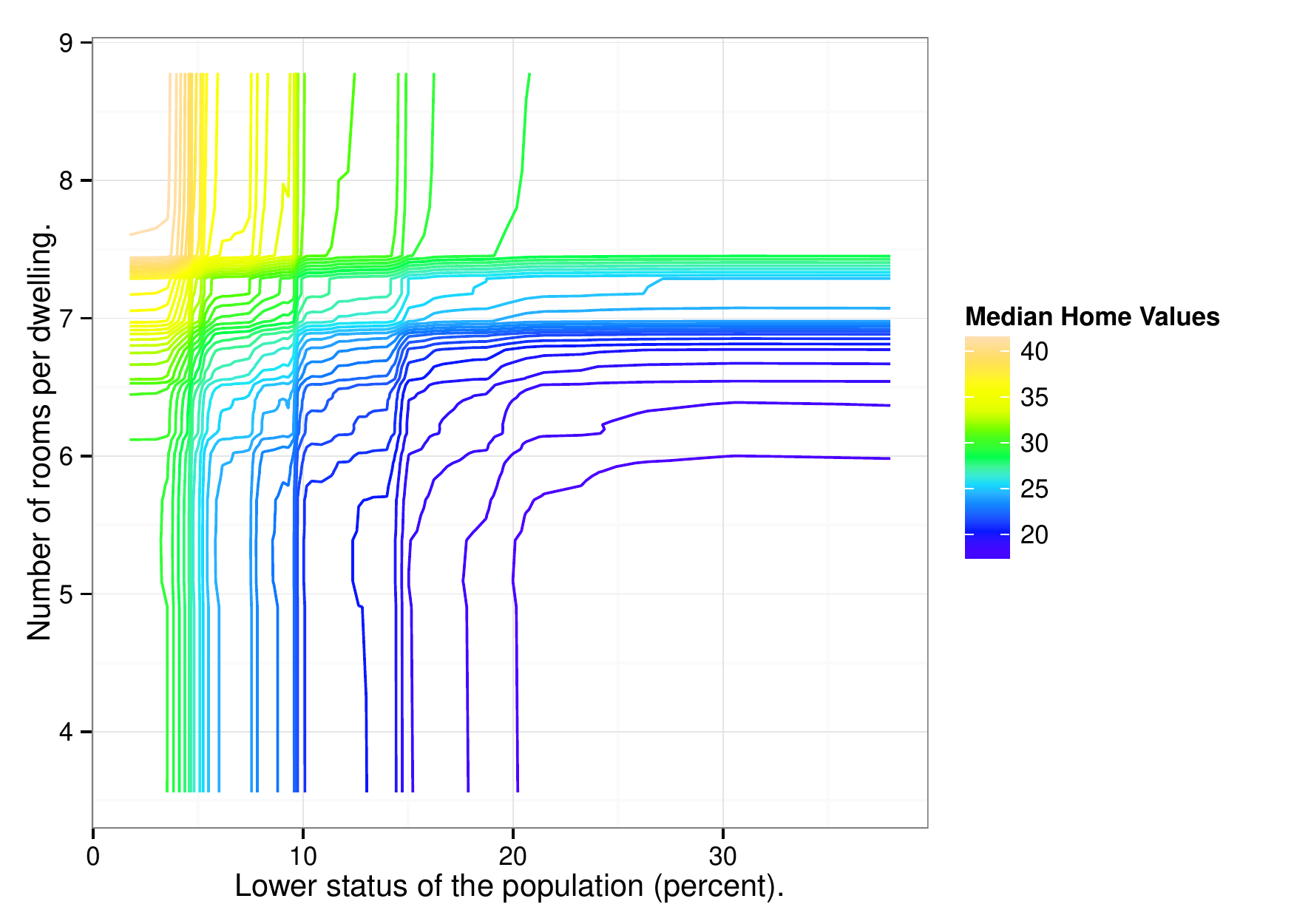} 

}

\caption[Partial coplot contour plot]{Partial coplot contour plot. Contours of median home value along the lstat/rm plane.}\label{fig:contour3d}
\end{figure}
\end{Schunk}

The contours are generated over the raw \code{gg_partial} estimation points, not smooth curves as shown in the  partial plot (Section~\ref{S:partialdependence}) and partial coplot (Section~\ref{S:partialcoplots}) figures previously. Contour lines, like topographic maps, are concentrated where the slope of the surface is large. We use color to indicate the direction of the contour lines, so that lower median home values are concentrated in the lower right hand corner, and the values increase along the diagonal toward the upper right. The close contour lines indicate some thing like a step in values at 7 and 7.5 rooms, and at 5, 10 and 15\% lstat.

Contour plots are still a little difficult to interpret. However, we can also generate a surface with this data using the \pkg{plot3D} package~\citep{plot3D:2014} and the \code{plot3D::surf3D} function. Viewed in 3D, a surface can help to better understand what the contour lines are showing us. 

\begin{Schunk}
\begin{Sinput}
R> # Modify the figure margins to make the figure larger
R> par(mai = c(0,0,0,0))
R> 
R> # Transform the gg_partial_coplot object into a list of three named matrices
R> # for surface plotting with plot3D::surf3D
R> srf <- surface_matrix(partial_surf, c("lstat", "rm", "yhat"))
R> 
R> # Generate the figure.
R> surf3D(x=srf$x, y=srf$y, z=srf$z, col=topo.colors(10),
+        colkey=FALSE, border = "black", bty="b2", 
+        shade = 0.5, expand = 0.5, 
+        lighting = TRUE, lphi = -50,
+        xlab="Lower Status", ylab="Average Rooms", zlab="Median Value"
+ )
\end{Sinput}
\begin{figure}[!htb]

{\centering \includegraphics[width=\maxwidth]{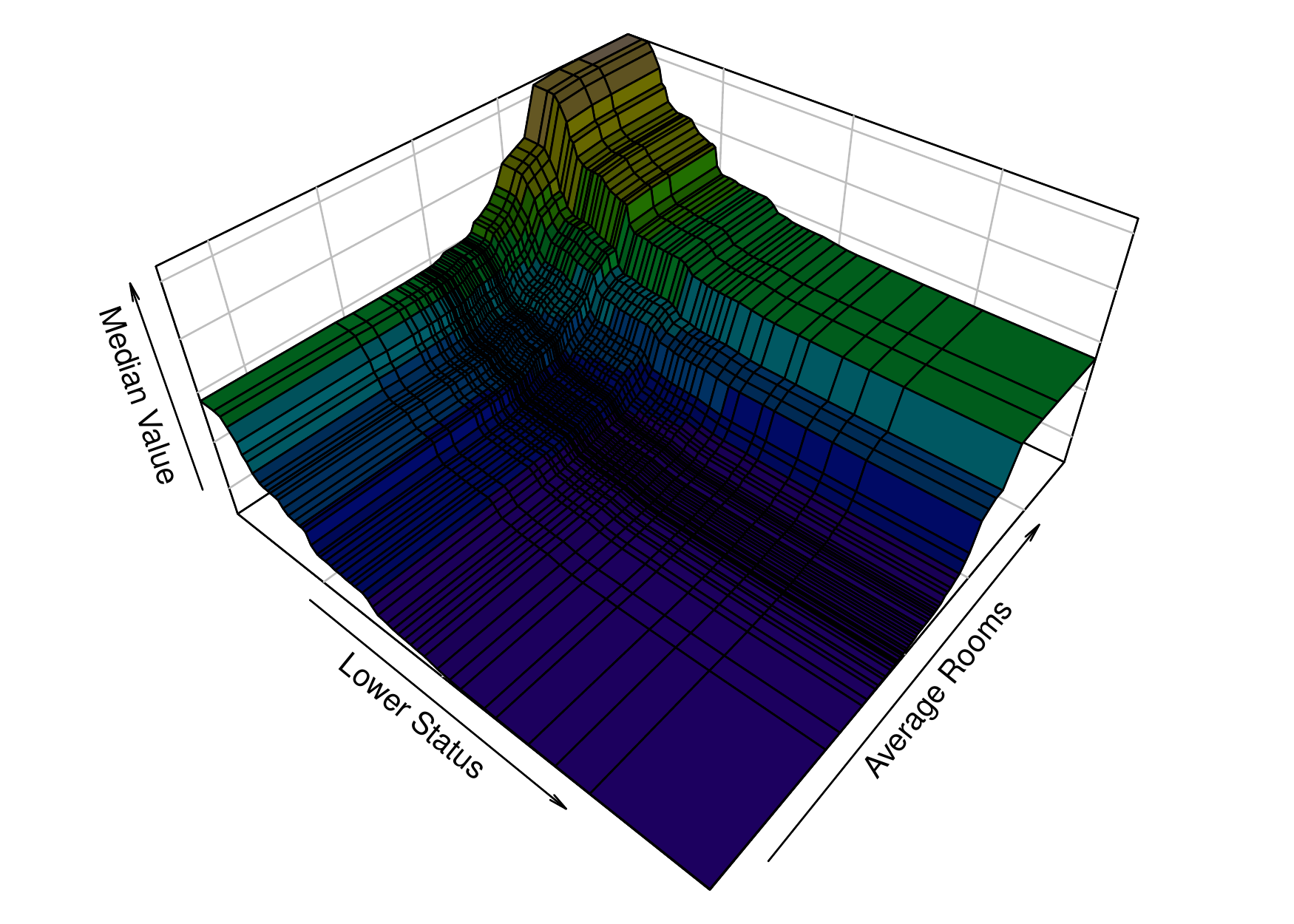} 

}

\caption[Partial plot surface]{Partial plot surface.}\label{fig:surface3d}
\end{figure}
\end{Schunk}

These figures reinforce the previous findings, where lower home values are associated with higher \code{lstat} percentage, and higher values are associated with larger \code{rm}. The difference in this figure is we can see how the predicted values change as we move around the map of \code{lstat} and \code{rm} combinations. We do still need to be careful though, as partial plots average over values on the surface that are note supported by actual observations.

\section{Conclusion} \label{S:conclusion}

In this vignette, we have demonstrated the use of the \pkg{ggRandomForests} package to explore a regression random forest built with the \pkg{randomForestSRC} package. We have shown how to create a random forest model (Section~\ref{S:rfsrc}) and determine which variables contribute to the forest prediction accuracy using both  VIMP (Section~\ref{S:vimp}) and Minimal Depth (Section~\ref{S:minimaldepth}) measures. We outlined how to investigate variable associations with the response variable using variable dependence (Section~\ref{S:variabledependence}) and the risk adjusted partial dependence (Section~\ref{S:partialdependence}) plots. We've also explored variable interactions by using pairwise minimal depth interactions (Section~\ref{S:interactions}) and directly viewed these interactions using variable dependence coplots (Section~\ref{S:coplots}) and partial dependence coplots (Section~\ref{S:partialcoplots}). Along the way, we've demonstrated the use of additional commands from the \pkg{ggplot2} package for modifying and customizing results from \pkg{ggRandomForests}.

\bibliography{ggRandomForests}

\appendix

\end{document}